%% file: main.tex
\documentclass[11pt]{article}

\usepackage[utf8]{inputenc} %
\usepackage[T1]{fontenc}    %
\usepackage[table]{xcolor}  %
\definecolor{forest}  {rgb}{0,.4,0} 
\definecolor{midnight}  {rgb}{0,0,.5} 
\usepackage[pdftex,plainpages=false,pdfpagelabels,colorlinks=true,urlcolor=midnight,citecolor=forest]{hyperref}
\usepackage{graphicx}
\usepackage{url}            %
\usepackage{booktabs}       %
\usepackage{amsfonts}       %
\usepackage{nicefrac}       %
\usepackage{xfrac}
\usepackage{microtype}      %
\usepackage{lipsum}		%
\usepackage{natbib}
\usepackage{doi}

\usepackage{multirow}
\usepackage[utf8]{inputenc} %
\usepackage[T1]{fontenc}    %
\usepackage{hyperref}       %
\usepackage{url}            %
\usepackage{booktabs}       %
\usepackage{amsfonts}       %
\usepackage{nicefrac}       %
\usepackage{microtype}      %
\usepackage{xcolor}         %
\usepackage[labelfont=bf]{caption}
\usepackage{subcaption}
\usepackage{mathtools}
\usepackage{svg}
\usepackage{listings}
\usepackage[linesnumbered,boxed,algoruled,noline,noend]{algorithm2e}
\usepackage{caption} %
\usepackage{pdflscape}
\let\oldnl\nl%
\newcommand{\nonl}{\renewcommand{\nl}{\let\nl\oldnl}}%

\SetCommentSty{mycommfont}

\DeclarePairedDelimiter{\norm}{\lVert}{\rVert}
\DeclarePairedDelimiter{\set}{\{}{\}}
\DeclareMathOperator* {\argmin} {argmin~}

\usepackage[capitalize,noabbrev]{cleveref}
\usepackage[letterpaper,top=2cm,bottom=2cm,left=3cm,right=3cm,marginparwidth=1.75cm]{geometry}
\usepackage{bm}

\usepackage{tikz}
\usepackage{pgfplots}
\usetikzlibrary{patterns,arrows,calc,math,decorations.pathreplacing,positioning,shapes,shapes.geometric,backgrounds}
\definecolor{bblue}{HTML}{4F81BD}
\definecolor{rred}{HTML}{C0504D}
\definecolor{ggreen}{HTML}{9BBB59}
\definecolor{ppurple}{HTML}{9F4C7C}
\definecolor{yyellow}{HTML}{B8BD4F}

\newcommand{\E}{\mathbb{E}}
\newcommand{\R}{\mathbb{R}}
\newcommand{\scat}{\mathrm{scat}}
\newcommand{\inn}{\mathrm{in}}

\newcommand{\lpf}{\mathsf{LPF}}
\newcommand{\cF}{\mathcal{F}}

\newcommand{\wideinputnet}{{MFISNet-Fused}}
\newcommand{\parallelnet}{{MFISNet-Parallel}}
\newcommand{\ournet}{{MFISNet-Refinement}}

\title{
Multi-Frequency Progressive Refinement for Learned Inverse Scattering}

\usepackage{authblk}

\author[1]{Owen Melia}
\author[1]{Olivia Tsang}
\author[2]{Vasileios Charisopoulos}
\author[3]{Yuehaw Khoo}
\author[3]{Jeremy Hoskins}
\author[1,2,3]{Rebecca Willett}
\affil[1]{Department of Computer Science, University of Chicago, US}
\affil[2]{Data Science Institute, University of Chicago, US}
\affil[3]{Computational and Applied Mathematics, Department of Statistics, University of Chicago, US}
\date{}                   
\begin{document}
\maketitle

\begin{abstract}
\input{revised_abstract}

\end{abstract}

\input{revised_intro}
\input{revised_setup_and_notation}

\input{revised_background}
\input{revised_our_method}
\input{revised_experiments}
\input{revised_conclusion}
\pagebreak

\input{revised_contribution}
\input{revised_data_availability}
\input{revised_acknowledgement}

\bibliographystyle{plainnat}
\bibliography{revised_reference}
\appendix
\input{revised_appendix_data_generation}

\input{revised_appendix_hyperparameter_optimization}
\input{appendix_warm_start_algo}
\input{revised_appendix_visualize_preds}

\end{document}

%% file: revised_abstract.tex
Interpreting scattered acoustic and electromagnetic wave patterns is a computational task that enables remote imaging in a number of important applications, including medical imaging, geophysical exploration, sonar and radar detection, and nondestructive testing of materials.
However, accurately and stably recovering an inhomogeneous medium from far-field scattered wave measurements is a computationally difficult problem, due to the nonlinear and non-local nature of the forward scattering process. 
We design a neural network, called Multi-Frequency Inverse Scattering Network (MFISNet), and a training method to approximate the inverse map from far-field scattered wave measurements at multiple frequencies. We consider three variants of MFISNet, with the strongest performing variant inspired by the recursive linearization method~\textemdash~a commonly used technique for stably inverting scattered wavefield data~\textemdash~that progressively refines the estimate with higher frequency content.
MFISNet outperforms past methods
in regimes with high-contrast, heterogeneous large objects, and inhomogeneous unknown backgrounds.

%% file: revised_intro.tex
\section{Introduction}

Wave scattering is an important imaging technology
with applications in medical and seismic imaging, sonar and radar detection, and nondestructive testing of materials.
In this setting, a known source transmits incident waves through a penetrable medium, and due to an inhomogeneity in the spatial region of interest, the incident waves are scattered. 
Several receivers measure the scattered wave field at distant locations. 
We are interested in the inverse wave scattering problem: given a set of scattered wave field measurements, we want to recover the inhomogeneity in the spatial region of interest that produced the measurements.
In this paper, we focus on the inverse wave scattering problem with unknown medium
and full-aperture measurements at multiple incident wave frequencies.
This problem is characterized by a highly nonlinear forward measurement operator, making the recovery of the scattering potential challenging.
We propose a machine learning solution to this problem: given a training set of pairs of scattering potentials and scattered wavefield measurements, we seek to approximate the inversion map with a deep neural network that predicts
a scattering potential from
scattered wavefield measurements at multiple frequencies.
We design a new training method and a new neural network architecture to achieve this goal.

The aforementioned inverse wave scattering problem has been widely studied. 
While the measurement operator is known to be injective when there are infinitely many sensors positioned in a ring around the scattering potential \citep{colton_looking_2018}, 
computational approaches must always operate in the ill-posed case where finite receivers are present. 
Thus, past research has focused on optimization approaches to solving the inverse problem.
Simple gradient-based optimization approaches to this problem face two major difficulties: computing a gradient requires solving a partial differential equation (PDE), which can be computationally expensive; additionally, the nonlinearity of the forward model induces a non-convex objective function.
Therefore, convergence of local search methods such as gradient descent is not guaranteed without careful initialization.

A classical strategy to alleviate the optimization challenges associated with a nonlinear forward model is to use a linear approximation. 
This linear approximation allows one to formulate the inverse problem as a linear least squares problem, which is relatively easy to solve.
A well-known method inverting this global linear approximation is the so-called \emph{filtered back-projection (FBP) method} \citep{natterer_mathematics_2001}. 
While this method is relatively easy to implement, it suffers from modeling errors and produces inaccurate reconstructions. 
To remedy this, many machine learning approaches are aimed at constructing data-driven approximations of the inverse map by designing architectures to imitate FBP~\citep{fan_solving_2019,khoo_switchnet_2019,li_wide-band_2022}.  
At a high level, these works approximate the two key components of FBP -- namely, an application of the adjoint of a linearization of the forward operator followed by a filtering step -- by suitably chosen neural network blocks.
As a result, they suffer from similar drawbacks as the FBP method: in particular, they provide low-quality reconstructions of high-contrast scattering potentials, especially in the presence of unknown inhomogeneous backgrounds or measurement noise.
A natural alternative is integrating machine-learning models into other iterative methods, which are computationally demanding relative to FBP but can provide higher quality reconstructions. %

A standard approach, which has been successful in the strongly nonlinear scattering regime, is to use data collected at multiple incoming wave frequencies. 
Recursive linearization methods \citep{chen_recursive_1995, bao_numerical_2003,borges_high_2016} use multi-frequency measurements to solve a sequence of sub-problems, starting at the lowest frequency to provide an initial estimate of the scattering potential and refining that estimate at progressively higher frequencies
using warm-started local search methods.
Algorithms in this family offer two benefits:
first, they alleviate the need for careful initialization since the loss landscape of the lowest frequency sub-problem
is typically well-behaved; second, they greatly reduce the number of PDE
solves by relying on first-order approximations of the forward model that are
relatively inexpensive to invert.
However, these methods require measurements at a large number of incident wave frequencies and still involve solving large-scale PDEs and least-squares problems for each frequency. 
This requires, for example, multiple CPU core-hours to recover a single image, even with a state-of-the-art PDE solver \citep{borges_high_2016}.

In light of these advances, we propose a new architecture and training method inspired by the recursive linearization algorithm. Our primary approach is based on a residual update architecture and training method that ensures specific network blocks solve specific sub-problems. 
In addition, we introduce two new methods of extending FBP-inspired neural networks to the multi-frequency setting. We call our networks ``Multi-Frequency Inverse Scattering Networks'', or ``MFISNets''.
We note that our methods are based on a neural network architecture from \cite{fan_solving_2019} but could, in principle, use any other neural network architectures designed for the inverse scattering problem in the single-frequency setting.

\subsection{Contributions \& paper outline}
In \cref{sec:problem_setup}, we formally define the inverse scattering problem and the machine learning objective.
We present standard results about inverse scattering and survey related work in~\cref{sec:background_and_related_work}.
In~\cref{sec:our_method}, we review the recursive linearization algorithm and introduce our method, \ournet.
Finally, we introduce our other two methods, \wideinputnet\ and \parallelnet, and present a numerical evaluation of our methods in~\cref{sec:experiments}. Our main contributions can be summarized as follows:
    \begin{enumerate}
        \item We introduce ``\ournet'', short for ``Multi-Frequency Inverse Scattering Network with Refinement'', a neural network architecture and training method that is inspired by recursive linearization algorithms \citep{chen_recursive_1995,bao_numerical_2003,borges_high_2016}. (\cref{sec:our_method})
        \item We show that our network achieves lower errors than single-frequency methods \citep{fan_solving_2019} and multi-frequency methods \citep{li_wide-band_2022} in the literature as well as the other newly-introduced MFISNets in a high-contrast, noiseless, full-aperture setting. (\cref{sec:experiments_noiseless})
        \item We demonstrate numerically that our method is robust to moderate measurement noise. (\cref{sec:experiments_with_noise})
        \item We consider alternative training strategies and find that \ournet\ is robust to the choice of training method, suggesting the majority of improvement is due to the architecture. (\cref{sec:experiments_ablation_study})
        \item We publicly release our code \url{https://github.com/meliao/mfisnets} and dataset \url{https://doi.org/10.5281/zenodo.14514353}.

    \end{enumerate}

%% file: revised_setup_and_notation.tex
\section{Problem Setup and Notation}
\label{sec:problem_setup}
\begin{figure}[h!]
    \centering
    \includegraphics[width=0.5\linewidth]{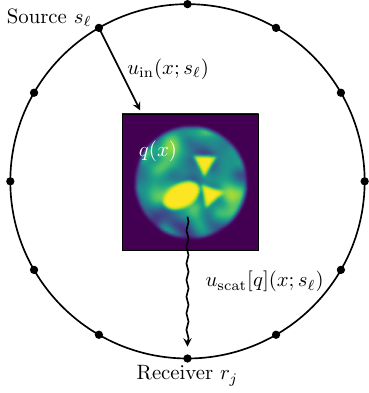}
    \caption{Geometry of the inverse scattering problem. An incident plane wave $u_{\inn}(x; s_\ell)$ coming from source direction $s_\ell$ interacts with the scattering potential $q(x)$. The resulting scattered wave field $u_{\scat}[q](x; s_\ell)$ is recorded by a receiver $r_j$.
    }
    \label{fig:diagram}
\end{figure}

The forward model for our imaging setup is implicitly defined by a PDE problem involving the Helmholtz equation. 
See \cref{fig:diagram} for a diagram of the geometry of the problem. 
Let $x\in \R^2$ be the spatial variable. 
Suppose $u_{\inn}(x; s) = e^{ i k x \cdot s }$ is an incoming plane wave with direction $s\in \mathbb{S}^1$, wavelength $\lambda$, and angular wavenumber $k=2\pi/\lambda$.
We normalize the problem's units so this wave travels at speed $c_0\equiv 1$ in free space.
The incoming wave interacts with a real-valued scattering potential $q(x)$ to produce an additive perturbation, called the scattered wave field $u_{\scat}[q](x;s)$.
We define $q(x)=c_0^2/c^2(x)-1$ where $c(x)$ is the wave speed at $x$. The total wave field $u[q](x;s) = u_{\scat}[q](x;s) + u_{\inn}(x; s)$ solves the following inhomogeneous Helmholtz equation, and the scattered wave field satisfies the Sommerfeld radiation boundary condition:
\begin{align}
    \begin{cases}
    \Delta u[q](x; s) + k^2 (1 + q(x)) u[q](x; s) = 0 & x \in \R^2; \\[6pt]
    \dfrac{\partial u_{\scat}[q](x; s)}{\partial \| x \|_2} - i k  u_{\scat}[q](x;s) = o( \| x \|_2^{-1/2}), & \text{as $\|x \|_2 \to \infty$.}
    \end{cases}
    \label{eq:pde_problem}
\end{align}
We assume $q(x)$ is supported on a square domain $\Omega = [-0.5, 0.5]^2$, and we work with $q \in \mathbb{R}^{N_q \times N_q}$, the discretization of $q(x)$ onto a regular grid with $(N_q, N_q)$ grid points. 
We place the receivers equally spaced around a large ring of radius $R \gg 1$, centered at the origin. We identify individual receivers by their unit-vector directions $r_j \in \mathbb{S}^1$. 
We compute the solution for the same set of $N_r$ receiver points and $N_s$ incoming source directions, where $N_r=N_s$ and the grid points are equally distributed about the unit circle.
This results in a set of $(N_{r}, N_{s})$ observations, $\{ u_{\scat}[q](R r_j; s_\ell)  \}_{j, \ell \in [N_r] \times [N_s]}$, which we arrange in a data array $d_k \in \mathbb{C}^{N_r \times N_s}$.
We call the mapping from $q$ to $d_k$ the forward model with incoming wave frequency $k$:
\begin{align}
    \left( d_{k} \right)_{j, \ell} = \mathcal{F}_k[q]_{j, \ell} \equiv u_{\scat}[q](R r_j; s_\ell)
    \label{eq:forward_model}
\end{align}

Because we are interested in multi-frequency algorithms, we are interested in observations of the forward model evaluated on the same $q$ but with a set of incoming wave frequencies $[k_1, \hdots , k_{N_k}]$. In particular, our goal is to
approximate the following mapping:
\begin{align}
     \left[ \mathcal{F}_{k_1}[q], \hdots,  \mathcal{F}_{k_{N_k}}[q] \right] \mapsto q.
     \label{eq:multi_freq_forward_map}
\end{align}
Our goal is to train a neural network $g_\theta$ with parameters $\theta$ to approximate the mapping: $g_\theta( \mathcal{F}_{k_1}[q], \hdots, \mathcal{F}_{k_{N_k}}[q]) \approx q$. Given a distribution $\mathcal{D}$ over scattering potentials $q$, we draw a training set of $n$ independent samples from $\mathcal{D}$ to generate data to train the neural network. After evaluating the forward model $n N_k$ times, we have a training set
\begin{align}
\mathcal{D}_{n} := \left\{
    (q^{(j)}, \mathcal{F}_{k_1}[q^{(j)}], \hdots, \mathcal{F}_{k_{N_k}}[q^{(j)}])
\right\}_{j=1}^n
\label{eq:dataset}
\end{align}
We evaluate networks in this setting by measuring the relative $\ell_2$ error:
\begin{align}
    \text{RelativeL2Error}(g_\theta) &= \E_{q \sim \mathcal{D}} \left[ \frac{ \norm[Big]{g_\theta \left( \mathcal{F}_{k_1}[q], \hdots, \mathcal{F}_{k_{N_k}}[q] \right) - q}_2 }{ \| q \|_2} \right]
    \label{eq:relative_l2_error}
\end{align}
In practice, we approximate the expected relative $\ell_2$ error in \eqref{eq:relative_l2_error} by an empirical mean over a held-out test set of 1,000 samples drawn independently from $\mathcal{D}$.

%% file: revised_background.tex
\section{Background and Related Work}
\label{sec:background_and_related_work}

\subsection{Background}

\label{sec:background}
In this section, we review standard results about $\cF_k$ relevant to our study. In particular, we focus on a linear approximation of $\cF_k$ that gives insights into the inverse scattering problem.

The first result is that $\cF_k$ becomes more nonlinear as the magnitude of the scatterer $\|q \|_2$ or the wavenumber $k$ increases.
Indeed, the solution to \eqref{eq:pde_problem} can be equivalently defined as the solution to the Lippmann-Schwinger integral equation: 
\begin{align}
  u_\scat[q](x;s) = k^2 \int_\Omega G_k(\|x-x' \|_2) q(x') (u_\inn(x';s)+u_\scat[q](x';s)) \text{d}x'
  \label{eq:lippmann-schwinger}
\end{align}
where $G_k$ is the Green's function for the homogeneous Helmholtz operator. 
This recursive equation provides a nonlinear map from $q(x)$ to $u_{\scat}[q](\cdot; s)$ and therefore $\mathcal F_k[q]$.

One way to view this nonlinearity is to interpret the Lippmann-Schwinger equation as a power series in $q(x)$ by iteratively substituting the value of $u_\scat[q](x)$ into its appearance on the right-hand side of \eqref{eq:lippmann-schwinger}.
For example, performing this substitution once yields a linear and a quadratic term in $q(x)$ that are independent of $u_{\scat}[q]$, as well as a ``remainder'' term involving the unknown $u_{\scat}[q]$ that accounts for higher-order terms:
\begin{align}
u_\text{scat}[q](x;s) &= k^2 \int_\Omega G_k(\| x-x'\|_2)q(x') u_\text{in}(x';s) \text{d}x' \nonumber \\
+& k^4 \int_\Omega G_k(\| x-x'\|_2)q(x') \int_\Omega G_k(\| x'-x''\|_2)q(x'') u_\text{in}(x'';s) \text{d}x'' \text{d}x' \nonumber \\
+& \text{Higher-order terms in $k$ and $q(x)$}.
\label{eq:lippmann-schwinger-expanded}
\end{align}
This power series does not converge for general $q(x)$, but it helps illustrate which parts of the problem drive the nonlinearity of the operator $\mathcal{F}_k$: 
as $\| q \|_2$ or $k$ grow, the size of these nonlinear terms will also grow, and as a result $\mathcal{F}_k$ becomes increasingly nonlinear.

The next result is that, under a linear approximation, the far-field measurements are diffraction-limited and can only capture frequency components of $q$ up to $2k$.
Equivalently, the measurements depend on $q$ to a spatial resolution of $\lambda/2$.
We consider the first-order \emph{Born approximation} \citep{born_principles_1999}, which approximates \eqref{eq:lippmann-schwinger} by dropping the $u_\scat[q](x';s)$ term from the right-hand side.
This is further simplified with an approximation of the Green's function in the far-field limit \citep{born_principles_1999}, yielding
\begin{align}
d_k(r, s) &\approx k^2
 \int_\Omega e^{-ik (r-s)\cdot x'} q(x') \text{d}x'.
 \label{eq:far-field-data-fourier-integral}
\end{align}
We will refer to this linear approximation of the map from $q(x)$ to $d_{k}(r,s)$ as $F_k$. Note that $F_k q$ is proportional to the Fourier Transform of $q$ evaluated at frequency vectors of the form $k(r-s)$. 
Since $r,s\in\mathbb{S}^1$ 
range over the unit circle, the frequency vectors $k(r-s)$ take on values throughout a disk with radius $2k$ centered at the origin.
Thus, evaluations of the linearized forward model, $F_k q$, 
only contain the low-frequency components of $q$, while high-frequency components of $q$ are in the kernel of the linearized forward model $F_k$ \citep {chen_recursive_1995}. 

Owing to its simplicity, \eqref{eq:far-field-data-fourier-integral} is often used as inspiration for the design of neural network architectures approximating the inverse map $d_k \mapsto q$. 
The networks emulate the FBP method~\citep{natterer_mathematics_2001},
which produces an estimate $\hat{q}$ of the
scattering potential $q$ as
\begin{align}
\hat q = (F_k^*F_k + \mu I)^{-1} F_{k}^* d_k,
\end{align}
where
$F_k^*$ is the adjoint of $F_k$ and $\mu$ is a regularization parameter that stabilizes the inversion of $F_k^* F_k$. 
The operator $(F_k^*F_k+\mu I)^{-1}$ can be implemented as a two-dimensional spatial convolution \citep{khoo_switchnet_2019,fan_solving_2019,li_wide-band_2022}, while novel network architectures have been proposed to emulate $F_k^* \in \mathbb{C}^{N_q^2 \times N_r N_s}$ in a parameter-efficient manner.
In particular, \cite{fan_solving_2019} and~\cite{zhang_solving_2023} suggest leveraging the rotational equivariance of the forward model to emulate $F_k^*$ with one-dimensional convolutions,
after applying a far-field scaling and an appropriate coordinate transformation in~\citep[Equation 6]{fan_solving_2019}).
We refer to the network described by~\citet{fan_solving_2019} as FYNet.

\subsection{Related Work}
Deep learning has revolutionized linear inverse problems in imaging, advancing methods for superresolution, inpainting, deblurring, and medical imaging.
Many of these advances stem from methods combining deep neural networks with optimization algorithms.
For example, the \emph{deep unrolling} paradigm \citep{monga_algorithm_2021} performs a fixed number of steps of an iterative algorithm and replaces certain operations with learnable mappings, which are parameterized by neural networks whose weights are learned from data. Components that remain fixed throughout training may reflect prior knowledge of problem parameters, such as explicit knowledge of the forward measurement model. In this setting, the network is usually trained end-to-end by minimizing the Euclidean distance between the network outputs and the true data.
Another paradigm, called \emph{plug-and-play denoising} \citep{venkatakrishnan_plug-and-play_2013}, suggests that general image denoisers can be used in place of proximal operators for regularization functions, an important subroutine in many optimization routines for linear inverse problems in imaging.
In this setting, neural network blocks are often trained to solve a different task, such as denoising corrupted signals, and then used inside the inversion algorithm.
\citet{ongie_deep_2020} provides a review of deep learning for inverse problems in imaging.

Several works in the wave scattering literature attempt to solve the inverse scattering problem by augmenting an optimization algorithm with components learned from data. At inference time, these methods require running an iterative optimization algorithm. 
\citet{kamilov_plug-and-play_2017} develops a plug-and-play algorithm for inverse scattering, and show that various off-the-shelf denoisers can be applied as proximal operators. 
\citet{zhao_deep_2023} use an encoder network paired with a network emulating the forward model and suggest optimizing a latent representation of the scattering potential using stochastic gradient descent.
When only phaseless measurements of the scattered wave field $u_\scat$ are available, \cite{deshmukh_unrolled_2022} propose a network unrolling proximal gradient descent, where the proximal operator is a neural network learned from data. 
For the inverse obstacle scattering problem in two dimensions, \cite{zhou_neural_2023} propose using a fully-connected neural network to warm-start a Gauss-Newton algorithm. 
\citet{ding_coupling_2022} train a neural network to approximately invert a forward scattering process depending on temporal data, and use this approximate inverse as a nonlinear preconditioner for a nonlinear least squares optimization routine. 
In concurrent work, \citet{zhang_back-projection_2024} use diffusion sampling to reconstruct scattering potentials and quantify uncertainty of the reconstructions.

Other methods propose to learn the inverse map directly from data. Recently, neural networks that are approximately invariant to discretization size have been proposed as methods of learning maps between general function spaces \citep{li_fourier_2021,lu_learning_2021} and these general-purpose networks have been applied to inverse scattering \citep{ong_integral_2022}. 
Other networks have been designed to invert the forward scattering model in particular; see~\cite{chen_review_2020} for a broad review of such approaches. 
In our work, we are particularly interested in FBP-inspired methods. 
The work of~\cite{khoo_switchnet_2019} leverages the approximate low-rank structure of scattering operators to design their SwitchNet network. 
\citet{fan_solving_2019} propose a data transformation that facilitates emulating the adjoint of the linearized forward model via 1D convolutions. 
In our work, we consider how to combine multiple such network blocks to invert the multi-frequency forward map in \eqref{eq:multi_freq_forward_map}. 
One way of combining these blocks is to learn each adjoint operator $F_{k}^*$ 
as a separate neural network block and combine data to jointly emulate the learned filtering operators $( F^*_{k} F_{k} + \mu I)^{-1}$.
This strategy is employed in \cite{zhang_solving_2023}, which is similar to our \parallelnet, but uses a different parameterization for the layers emulating $F_{k}^*$ and   $( F^*_{k} F_{k} + \mu I)^{-1}$.
Another strategy is to use the Wide-Band Butterfly Network \citep{li_wide-band_2022,li_accurate_2021}, which hierarchically merges information from different frequencies in the network block emulating $F^*_{k}$. We provide more detail and commentary about methods of combining network blocks to form multi-frequency networks in \cref{sec:baselines}.

Finally, we note in passing that the design of our method is inspired by~\emph{homotopy methods}~\citep{watson_modern_1989}. These methods solve a sequence of sub-problems of increasing difficulty, gradually transforming a simple (but uninformative) optimization problem to the optimization problem of interest and using
solutions to a given sub-problem to warm-start local search methods for subsequent sub-problems.
Such a sequence can be constructed explicitly (\emph{e.g.}, by varying regularization levels) or implicitly; for example,
\emph{curriculum learning}~\citep{benigo_curriculum_2009} progressively adjusts the training data distribution from ``easy'' to ``hard'' samples and has been used to train physics-informed neural networks in challenging problem settings~\citep{krishnapriyan_characterizing_2021,huang_hompinns_2022}.

%% file: revised_our_method.tex
\section{Recursive Linearization and Our Method }
\label{sec:our_method}
We propose a neural network that learns to approximate the multi-frequency inversion map from training data. To design the network and training algorithm, we draw inspiration from the recursive linearization method for inverse scattering, which we briefly review below.

\subsection {Recursive Linearization}
\label{sec:recursive-linearization}
Recursive linearization is a classical method for solving the inverse scattering problem, introduced by \cite{chen_recursive_1995}. In spite of the nonlinearity of the true forward scattering model described in \eqref{eq:lippmann-schwinger}, recursive linearization breaks the inverse problem into a series of simpler problems, each of which corresponds to a linear inverse problem. In this section we discuss the intuition behind this strategy.

Recall from our discussion in~\cref{sec:background} that the forward map evaluated at low incident wave frequencies $k$ acts approximately like a low-pass filter with cutoff frequency $2k$. 
At first glance, this suggests that observing $\cF_k[q]$ for a high value of $k$ is sufficient for high-resolution recovery of $q$. 
However, when viewed from an optimization perspective, it becomes clear that this problem is increasingly challenging for large values of $k$. For example, one might consider the nonlinear least-squares problem
\begin{align}
    \underset{\hat q}{\text{argmin~}} \lVert d_k - \cF_k[\hat q] \rVert_2^2.
\label{eq:opt-program}
\end{align}

\begin{figure}[ht]
    \centering
    \includegraphics[width=0.95\linewidth]{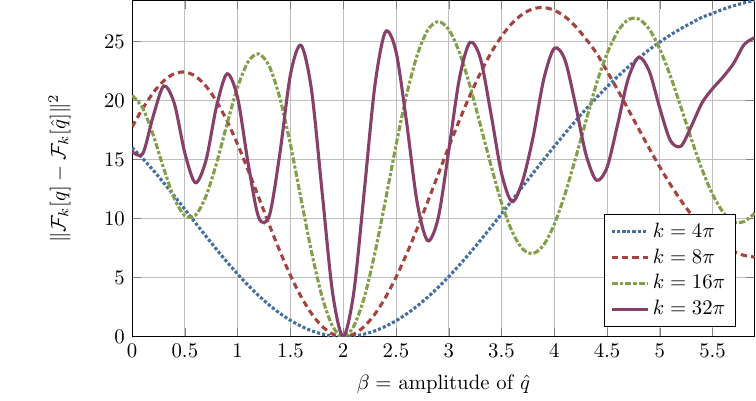}
    \caption{Even in a highly stylized setting, accurately and reliably inverting $\cF_k$ is difficult for high frequencies $k$.
    Suppose the ground-truth scattering potential is $q(x) = \beta \exp({- \frac{\|x \|^2 }{2 \sigma^2}})$, a Gaussian bump with known spread parameter $\sigma = 0.1$ but unknown amplitude $\beta$.
    We show the optimization landscape that arises from searching over different amplitudes. At low incident wave frequencies, this optimization landscape is smooth and has a large basin of attraction. However, as the incident wave frequency increases, the optimization landscape becomes highly oscillatory, requiring a nearly exact initialization to guarantee convergence to the ground truth.
    In experimental settings, the parameterization of $q$ is often high-dimensional, which requires higher frequency data to resolve high-frequency information in $q$.
    This numerical example was inspired by \cite{bao_numerical_2003}.
    }
    \label{fig:low_dim_opt}
\end{figure}

To illustrate the challenges for the optimization formulation with increasing values of $k$, we consider a simple example where $q$ is known to be a Gaussian bump with a given spread parameter and unknown amplitude.
Given observations $d_k=\mathcal F_k[q]$ and a numerical PDE solver to calculate $\mathcal F_k[\cdot]$, one could estimate the amplitude of $q$ by solving a minimization problem similar to \eqref{eq:opt-program}, but only searching over the unknown amplitude.
In \cref{fig:low_dim_opt}, we plot this objective as a function of the amplitude of $\hat{q}$ for a range of values of $k$.
For large values of $k$, the objective function is highly oscillatory and contains many spurious local minima. 
Typically this suggests that it can be challenging to locate the global minimum unless there is a scheme to initialize estimates close enough to the global minimum to avoid getting stuck elsewhere.
However, \cref{fig:low_dim_opt} also shows that the objective function oscillates much more slowly at the smallest value of $k$ while sharing the same global minimum. This suggests that the low-frequency observations can be used to get close to the global minimum, even though they are diffraction-limited and cannot resolve high-frequency components.

The recursive linearization algorithm leverages this insight by solving a sequence of inversion problems at increasing wave frequencies $k$. Crucially, each sub-problem uses the output of the previous sub-problem to initialize a new optimization problem. This method was introduced in \cite{chen_recursive_1995} and further developed in \cite{bao_numerical_2003,borges_high_2016}. At iteration $t$, the algorithm uses the previous estimate $\hat{q}_{k_{t-1}}$ along with a new set of observations $d_{k_t}$, and it calculates an update $\delta q$ that minimizes the $\ell_2$ distance in measurement space:
\begin{align}
     \argmin\limits_{\delta q} \| d_{k_t} - \cF_{k_t}[\hat{q}_{k_{t-1}} + \delta q] \|^2_2
    \label{eq:update_objective_1}
\end{align}
The value of $\hat q_{k_{t-1}}$ may make it possible to avoid spurious local minima, but this problem is still difficult since an iterative optimizer would require solving a PDE at each of its iterations.
However, $\cF_{k_t}[\hat{q}_{k_{t-1}} + \delta q]$ is well-approximated by a first-order Taylor expansion about $\hat q _{k_{t-1}}$ when $\delta q$ is small or when it does not contain low-frequency information \citep{chen_recursive_1995}. This motivates the following surrogate for the optimization problem in~\eqref{eq:update_objective_1}:
\begin{align}
    \argmin\limits_{\delta q} \| d_{k_t} - \left( \cF_{k_t}[\hat{q}_{k_{t-1}}] +  D\cF_{k_t}[\hat{q}_{k_{t-1}}] \delta q \right) \|^2_2,
    \label{eq:update_objective_2}
\end{align}
where $D\cF_{k_t}[\hat{q}_{k_{t-1}}]$ denotes the Fr\'echet derivative of the forward model at $\hat{q}_{k_{t-1}}$. 
The action of $D\cF_{k_t}[\hat{q}_{k_{t-1}}]$ and its adjoint, $D\cF_{k_t}^*[\hat{q}_{k_{t-1}}]$, can be computed using the adjoint-state method \citep{bao_numerical_2003,borges_high_2016}.
The resulting algorithm is akin to a Gauss-Newton method; critically, each sub-problem of the form shown in~\eqref{eq:update_objective_2} is a linear least-squares problem.
We outline a sketch of the recursive linearization algorithm in \cref{alg:recursive-linearization}. 

The recursive linearization algorithm is very demanding computationally. 
Each iteration requires solving $N_s$ large-scale PDEs and a high-dimensional least-squares problem, which quickly creates a large computational burden when producing high-resolution solutions. 
In a classical setting without machine learning, the frequencies should be spaced close to each other for best results. \citet{chen_inverse_1997} uses $k=1,2,\dots,9$ in their numerical experiments, while \cite{borges_high_2016} uses $k=1,1.25,\dots,70$, which they report takes around 40-50 hours per sample to produce a single $241 \times 241$ pixel image.

\begin{algorithm}[h!]
    \caption{
        Recursive Linearization for Inverse Scattering based on \cite{chen_recursive_1995,chen_inverse_1997,bao_numerical_2003,borges_high_2016}
    }
    \label{alg:recursive-linearization}
    \KwIn{Multi-frequency data $\{d_{k_1}, d_{k_2}, \dots, d_{k_{N_k}}\}$}
    $\hat {q} _{k_1} \gets  \left( F_{k_1}^* F_{k_1} + \mu I \right)^{-1} F_{k_1}^* d_{k_1}$ \\
    \For(){$t=2,\dots,N_k$} {
        Compute $\cF_{k_t}[\hat{q}_{k_{t-1}}]$ and $D\cF_{k_t}[\hat{q}_{k_{t-1}}]$ \label{alg:rl-inner-loop-1}\\
        $\delta q_{k_t} \gets \argmin\limits_{\delta q} \| d_{k_t} - \left( \cF_{k_t}[\hat{q}_{k_{t-1}}] +  D\cF_{k_t}[\hat{q}_{k_{t-1}}] \delta q \right) \|^2_2$ \label{alg:rl-inner-loop-2}\\
        $\hat q_{k_t} \gets \hat q_{k_{t-1}} + \delta q_{k_t}$ \\
    }
    \KwResult {Final estimate $\hat q_{k_{N_k}}$}
\end{algorithm}

Although recursive linearization is computationally expensive as stated, we believe that one of the key features of recursive linearization is the way that it breaks the recovery process into multiple steps, each of which refines the estimate from the previous step using data of a higher frequency.
We will refer to this step-wise recovery strategy as progressive refinement.

Progressive refinement facilitates the recovery process, since each step is only responsible for a correction to the estimate of the scattering potential within a frequency band.
Focusing on this strategy also allows us to look for machine learning methods that do not explicitly emulate $\mathcal F_k[\cdot]$ or $D\mathcal F_k[\cdot]$, which are expensive to compute.

To this end, we consider a generalization of recursive linearization where we replace lines \ref{alg:rl-inner-loop-1} and \ref{alg:rl-inner-loop-2} in \cref{alg:recursive-linearization} by describing the inner loop as a generic refinement step:
\begin{align}
    \delta q_{k_t} 
    = \text{RefinementStep}_{k_t}(\hat q_{k_{t-1}}, d_{k_t}) \qquad \text{for $t=2,\dots,N_k$}
    \label{eq:iterative-refinement-recurrence}
\end{align}

\noindent where $\text{RefinementStep}_{k_t}(\hat q_{k_{t-1}}, d_{k_t})$ refers to the update calculated for estimate $\hat q_{k_{t}}$ given data $d_{k_t}$ and can be implemented using a neural network. We will propose and discuss a network architecture in the next section.

\subsection {Our Method}
We use~\cref{alg:recursive-linearization} as inspiration for the design of our neural network architecture and training method. In particular, we focus on the following two crucial aspects of~\cref{alg:recursive-linearization}:
\begin{description}
    \item[Progressive refinement:] The algorithm builds intermediate estimates of the scattering potential which are progressively refined with the introduction of new data.
    \label{item:aspect-1}
    \item[Homotopy through frequency:] The iterative refinements from the first step form a homotopy from low to high-frequency measurements. As a result, updates at step $t$ contain high-frequency information relative to $k_{t-1}$.
\end{description}

\begin{figure}
    \begin{subfigure}[b]{0.5\textwidth}
        \centering
        \includegraphics[width=0.95\linewidth]{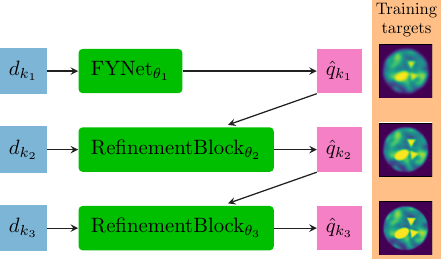}
        \caption{\ournet~Architecture}
        \label{fig:architecture}
    \end{subfigure}
    \begin{subfigure}[b]{0.5\textwidth}
        \centering
        \includegraphics[width=0.95\linewidth]{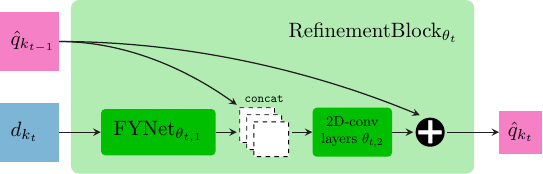}
        \caption{Refinement Block}
        \label{fig:update_block}
    \end{subfigure}
    \caption{Our \ournet~architecture is designed to emulate the recursive linearization algorithm. \cref{fig:architecture} shows that our network proceeds by making an initial low-frequency reconstruction and then making a series of updates given higher-frequency data and an estimate of the scattering potential. The network is trained to match the intermediate reconstructions to low-pass filtered versions of scattering potentials from the training set. One example collection of such filtered scattering potentials is shown. \cref{fig:update_block} shows that our refinement block is a simple extension of the FYNet architecture from \cite{fan_solving_2019}. By using a skip connection in this block, we ensure the network only needs to predict an update to the estimated scattering potential.
    }
    \label{fig:architecture_all}
\end{figure}

To emulate the progressive refinement structure, we propose a network with a residual structure and skip connections. The network comprises multiple blocks, one for each
incident wave frequency $k_t$, $t = 1, \dots, N_{k}$.
The input to each block is measurement data $d_{k_t}$ collected at a particular incident wave frequency ${k_t}$.
The input passes through an FYNet block  \citep{fan_solving_2019}, which approximately inverts the forward model.
The output of the FYNet block is then concatenated with the output of the previous block, $\hat{q}_{k_{t-1}}$, and the concatenation is passed to 2D convolutional layers for a second filtering step. 
Finally, a skip connection adds $\hat{q}_{k_{t-1}}$ to the output of the last convolutional layer of the block, producing
the next estimate $\hat{q}_{k_t}$.
The network's architecture is shown in \cref{fig:architecture_all};
we call the resulting network ``\ournet.''
Note that, under this construction, the FYNet blocks could be replaced by any other neural network architecture designed for the single-frequency inverse scattering problem.

To emulate the homotopy through frequency, we design a training method to ensure each successive block adds higher-frequency information to the estimate of the scattering potential.
Under the Born approximation~\eqref{eq:far-field-data-fourier-integral}, we know $d_{k}$ contains information about $q$ up to frequency limit $2k$. 
This suggests that given data $d_{k_t}$, we should be able to reconstruct the frequency components of $q$ up to $2k_t$. 
To reflect this, we train the output of block $t$ with the following loss function:
\begin{align}
    L_t(\hat{q}_{k_{t}}; q) &=
    \mathbb{E}_{q \sim \mathcal{D}_{n}} \left[ \| \hat{q}_{k_{t}} - \lpf_{2k_t} q \|^2 \right]
    \label{eq:loss_i}
\end{align}
In~\eqref{eq:loss_i}, $\lpf_{2k_t}$ is a low-pass filter with approximate cutoff frequency $2k_t$, implemented as a Gaussian filter to avoid ringing artifacts; its frequency response is given in the Fourier domain by
\begin{align}
    \widetilde{\lpf}_{f_{\mathsf{cut}}}(f) := \exp\left(
        -\frac 1 2 \left(\frac{f}{f_{\mathsf{cut}} / \sqrt{2\log 2}} \right)^2
    \right),
    \label{eq:filter-freq-response}
\end{align}
given a target cutoff frequency $f_{\mathsf{cut}}$.
Note that we shrink the standard deviation by a factor of $\sqrt{2\log 2}$ so that the filter's half-width at half-maximum matches $f_{\mathsf{cut}}$.

To train the network, we first adjust the weights of each block in a sequential fashion and then perform a final training step which fine-tunes all of the blocks jointly; the training procedure is summarized in \cref{alg:training_procedure}.

\begin{algorithm}[h!]
\DontPrintSemicolon
    \KwIn{Randomly-initialized neural network parameters $\{\theta_1, \dots, \theta_{N_{k}} \}$;\newline
         Training data samples $\mathcal{D}_{n} := \left\{
            (q^{(j)}, d_{k_1}^{(j)}, \hdots, d_{k_{N_{k}}}^{(j)})
        \right\}_{j=1}^n$.}
    \For(){$t=1,...,N_{k}$}{
        Set $\theta_t$ as trainable, and freeze all other weights \\
        \If{$t < N_k$}{
        Train $\theta_t$ by optimizing $L_t$ \tcp*{\cref{eq:loss_i}}
        }
        \Else{
        Train $\theta_t$ by optimizing $\| \hat{q}_{k_{N_{k}}} - q \|_2^2$
        }
    }
    Set all weights as trainable \\
    Train all weights by optimizing $\| \hat{q}_{k_{N_{k}}} - q \|_2^2$ \\
    \KwResult{Trained neural network parameters $\{\theta_1, ..., \theta_{N_{k}} \}$.}
    \caption{Training Procedure}
    \label{alg:training_procedure}
\end{algorithm}

\subsection{Implementation of FYNet}
We implement the inversion network described in \citep{fan_solving_2019} and call it FYNet. This method suggests applying the far-field scaling and a transformation from the receiver and source direction coordinates $(r,s)$ to a new set of variables
as summarized in~\citep[Equation (6)]{fan_solving_2019},
which we perform using bicubic interpolation. We describe this transformation in \cref{eq:mh_transform}. 
We split the complex-valued input into real and imaginary parts along a channel dimension.
To implement the action of the adjoint operator $F_k^*$, we use a composition of three 1-dimensional convolutional layers. 
We implement the convolutional layers as learnable in the Fourier domain because the expression for $F_k^*$ derived in \citep{fan_solving_2019} is local in frequency, but not space. 
To implement the filtering operator $\left( F_k^* F_k + \mu I \right)^{-1}$, we use a composition of three 2-dimensional convolutional layers. All layers but the last one use ReLU activations. 
This network outputs an estimate of the scattering potential on a regular polar grid, in coordinates $(\rho, \phi)$. 
Following \citep{fan_solving_2019}, we train the network by minimizing the difference between predictions and targets on the polar grid. 
We transform to Cartesian coordinates for visualization and computing final test statistics. 

%% file: revised_experiments.tex
\section{Experiments}
\label{sec:experiments}
In this section, we describe the setting and results for our numerical investigation of the efficacy of our proposed method.

\subsection{Dataset and Data Generation}
\label{sec:dataset}

\begin{figure}[!ht]
    \begin{subfigure}[b]{0.33\textwidth}
        \centering
        \includegraphics[width=\linewidth]{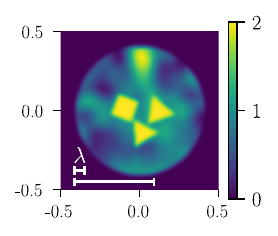}
        \caption{Scattering potential $q$}
        \label{fig:q}
    \end{subfigure}%
    \begin{subfigure}[b]{0.33\textwidth}
        \centering
        \includegraphics[width=\linewidth]{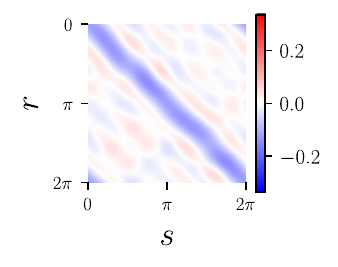}
        \caption{Measured $d_k$, $k=4\pi$}
        \label{fig:d_rs_nu_2}
    \end{subfigure}%
        \begin{subfigure}[b]{0.33\textwidth}
        \centering
        \includegraphics[width=\linewidth]{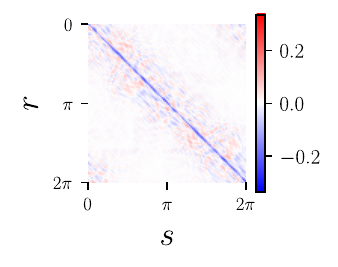}
        \caption{Measured $d_k$, $k=32\pi$}
        \label{fig:d_rs_nu_16}
    \end{subfigure}
    \caption{\cref{fig:q} shows a typical example from our distribution of scattering potentials, drawn from the test set. Our distribution of scattering potentials has a random low-frequency background field, occluded by piecewise constant geometric shapes. The bottom-left corner shows the wavelength of two incident waves with frequencies $k=4\pi$ and $k=32\pi$. 
    \cref{fig:d_rs_nu_2,fig:d_rs_nu_16} show the output of the forward model applied to this scattering potential, using these incident frequencies. The real part of $u_{\scat}$ is shown.
    }
    
    \label{fig:target_and_preds.}
\end{figure}
\paragraph{Distribution of Scattering Potentials} We define a distribution of scattering potentials $\mathcal{D}$ which has nonzero spatial support on the disk of radius $0.4$, with a smoothly varying random background occluded by three randomly placed and randomly sized piecewise-constant shapes. 
We normalize the scattering potential so the background has minimum and maximum values $0$ and $2$ respectively, and we normalize the piecewise-constant shapes to have value $2$. \cref{fig:q} shows one such scattering potential from our distribution.

Notably, the contrast of these scattering potentials $\| q \|_\infty =2$, which is much larger than the contrast used in distributions to evaluate other machine learning methods in the shape reconstruction regime \cite{fan_solving_2019, li_wide-band_2022}.
The high-contrast regime is an important experimental setting because it ensures the nonlinearity of the forward model, which is the difficult and interesting problem setting, is captured.
The non-constant background also adds to the difficulty of the problem by increasing $\| q \|_2$, which adds to the nonlinearity of the forward model. It also adds much more entropy to $\mathcal{D}$. 
We use this model to reflect experimental conditions in imaging tasks, wherein backgrounds are rarely known, constant, or homogeneous.

\label{sec:data-gen}
\paragraph{Implementation of $\mathbf{\mathcal{F}}$} To implement the forward model, we implement a numerical PDE solver to compute solutions of \eqref{eq:pde_problem}. 
We implement this by discretizing the scattering domain $\Omega$ with a $(N_q, N_q)$ regular grid, with $N_q=192$. 
We transform \eqref{eq:pde_problem} into the Lippmann-Schwinger integral equation and recast the latter as a sparse linear system, 
which we accelerate using fast Fourier transforms and hardware acceleration. We solve this linear system with GMRES~\citep{saad_gmres_1986} implemented by SciPy \citep{virtanen_scipy_2020} to a relative tolerance of $10^{-2}$.
This formulation allows us to compute the solution $u_{\scat}$ on a large, distant ring placed at radius $R=100$. 
We compute the solution at $N_r = 192$ equally-spaced positions on this ring, and we repeat this process for each of the $N_s=192$ equally-spaced source directions. The sources and receivers are located on the same grid.
The runtime for evaluating the full forward model for
a given $q$ at the lowest and highest incident wave frequencies considered requires 5 and 220 seconds respectively, using one $\text{NVIDIA}^{\text{\textregistered}}$ A40 GPU. \cref{fig:d_rs_nu_2,fig:d_rs_nu_16} show the solution $u_\scat$ produced by our implementation for low and high incident wave frequencies, respectively.

\begin{sloppypar}
\citet{fan_solving_2019} use a coordinate transformation of the far-field scattering data from coordinates $(r, s)$ to $(m, h)$, where
\end{sloppypar}
\begin{align}
    m := \frac{r+s}{2} \quad\text{and}\quad h:= \frac{r-s}{2}
    \label{eq:mh_transform}
\end{align}
\begin{sloppypar}
    \noindent
as described in \citep[Equation (6)]{fan_solving_2019}. The variable $m$ ranges from $0$ to $2\pi$, while $h$ ranges from $-\pi/2$ to $\pi/2$. As in \cite{fan_solving_2019}, we choose $N_m=192$ and $N_h=96$ to keep the angular sampling frequency fixed for all angular variables considered in the problem. We perform this transformation using bicubic interpolation, and the resulting grid has dimensions $(N_m, N_h)=(192,96)$.
    
\end{sloppypar}

\begin{sloppypar}
The FYNet blocks reconstruct images on a regular polar grid with $(N_\rho, N_\phi) = (96, 192)$ pixels, and the radial dimension of our polar grid extends to $\rho_{\max} = 0.5$. Finally, we use bicubic interpolation to transform the model's outputs to the $(N_q, N_q)$ Cartesian grid for final visualization and error measurement.
\end{sloppypar}

\subsection{Alternative Multi-Frequency Methods}
\label{sec:baselines}
\begin{figure}
  \begin{subfigure}[b]{\textwidth}
      \centering
      \begin{tikzpicture}[node distance=1em, inner sep=1pt, square/.style={regular polygon,regular polygon sides=4}]

        \definecolor{hblue}{HTML}{209ca1}
        \definecolor{mblue}{HTML}{519cc8}
        \definecolor{lblue}{HTML}{908cc0}
        \definecolor{darkgray176}{RGB}{176,176,176}
        \definecolor{darkorange25512714}{RGB}{255,127,14}
        \definecolor{green01270}{RGB}{0,127,0}
        \definecolor{steelblue31119180}{RGB}{31,119,180}
    
        \node[fill=mblue!75!white, square] at (0, 0) (d1) {$d_{k_1}$};
        \node[fill=mblue!75!white, below=1em of d1, square] (d2) {$d_{k_2}$};
        \node[fill=mblue!75!white, below=1em of d2, square] (d3) {$d_{k_3}$};

        \draw[fill=green!75!black!30!white, draw=none, rounded corners=4pt] (d1)++(2em,2em) rectangle ([xshift=22.75em, yshift=-2em]d3.east);

        \node[
            right=1.5em of d1, inner sep=6pt, rounded corners=2pt, fill=green!75!black,
            align=center, scale=0.7] (v2) {$V^2$};
    
        \node[
            right=2.5em of v2, inner sep=6pt, rounded corners=2pt, fill=green!75!black,
            align=center, scale=0.7] (h2) {$H^2$};
        \node[
            right=1.5em of d2, inner sep=6pt, rounded corners=2pt,
            fill=green!75!black, align=center, scale=0.7] (v3) {$V^3$};
        \node[right=1em of v3, inner sep=6pt, rounded corners=2pt, fill=green!75!black,
            align=center, scale=0.7] (h3) {$H^3$};
        \coordinate[right=0.5em of h3] (e3);
    
        \node[right=1.5em of d3, inner sep=6pt, rounded corners=2pt, fill=green!75!black,
            align=center, scale=0.7] (v4) {$V^4$};
        \coordinate[right=1.5em of v4] (e4);
    
        \node[right=1.5em of d3, inner sep=6pt, rounded corners=2pt, fill=green!75!black,
            align=center, scale=0.7] (v4) {$V^4$};

        \draw[->, >=stealth, color=gray!25!black, semithick] (d1.east) -- (v2.west);
        \draw[->, >=stealth, color=gray!25!black, semithick] (d2.east) -- (v3.west);
        \draw[->, >=stealth, color=gray!25!black, semithick] (d3.east) -- (v4.west);
        \draw[->, >=stealth, color=gray!25!black, semithick]
        (v4.east) -| (h3.south);
        \draw[->, >=stealth, color=gray!25!black, semithick] (v3.east) -- (h3.west);
        \draw[->, >=stealth, color=gray!25!black, semithick]
        (h3.east) -| ([xshift=-0.2em]h2.south);
        \draw[->, >=stealth, color=gray!25!black, semithick] (v2.east) -- (h2.west);
    
        \coordinate[right=1.5em of h3] (e5);
    
        \node[
            right=2.5em of h3, inner sep=6pt, rounded corners=2pt, fill=green!75!black,
            align=center, scale=0.7] (switchnet) {Switch\\Layer};
        \draw[->, >=stealth, color=gray!25!black, semithick] ([xshift=0.2em, yshift=-0.05em]h2.south) |- (switchnet.west);
    
        \node[
            right=1.5em of switchnet, inner sep=6pt, rounded corners=2pt, fill=green!75!black,
            align=center, scale=0.7] (resnet) {ResNet\\Layers};
        \draw[->, >=stealth, color=gray!25!black, semithick] (switchnet.east) -- (resnet.west);
        
        \node[
            right=1.5em of resnet, inner sep=6pt, rounded corners=2pt, fill=green!75!black,
            align=center, scale=0.7] (upsamp) {Upsampling\\Layers};
        \draw[->, >=stealth, color=gray!25!black, semithick] (resnet.east) -- (upsamp.west);
    
        \node[above=2em of upsamp, xshift=-4em] (label) {Emulating $F^*_{k_t}$, $t=1,2,3$};

        \node[
            right=1.5em of upsamp, inner sep=6pt, rounded corners=2pt, fill=green!75!black,
            align=center, scale=0.7] (cnn) {
                2D-convs emulating\\%
                $\left(F_{k_t}^* F_{k_t} + \mu I\right)^{-1}$, $t=1,2,3$};
        \draw[->, >=stealth, color=gray!25!black, semithick] (upsamp.east) -- (cnn.west);

        \node[right=1.25em of cnn, fill=magenta!50!white, square, inner sep=1pt] (output) {${\hat{q}}$};
        \draw[->, >=stealth, semithick] (cnn.east) -- (output.west);

    \end{tikzpicture}
      \caption{Wide-Band Butterfly Network}
      \label{fig:architecture_widebnet}
  \end{subfigure}\\[2em]
  \begin{subfigure}[b]{0.64\textwidth}
      \centering
      \includegraphics[height=6.4em]{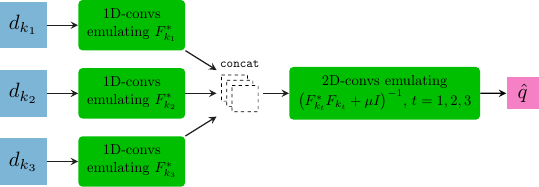}
      \caption{\parallelnet}
      \label{fig:architecture_parallelfynet}
  \end{subfigure}~
  \begin{subfigure}[b]{0.32\textwidth}
      \centering
      \includegraphics[height=6.4em]{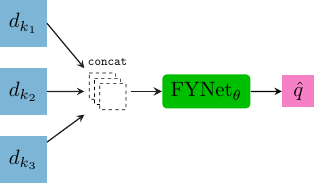}
      \caption{\wideinputnet}
      \label{fig:architecture_wide_input_fynet}
  \end{subfigure}

  \caption{Alternative neural network architectures for learning the multi-frequency inverse map. All blocks in dark green contain trainable parameters.}
  \label{fig:architecture_baselines}
\end{figure}

We wish to compare our method, \ournet, with other methods of learning an inverse to the multi-frequency forward map. 
We design two new methods of extending FBP-inspired single-frequency architectures to the multi-frequency setting. 
We use the FYNet architecture \citep{fan_solving_2019} to instantiate all three of our MFISNet methods, which allows us to focus on the effects caused by different methods of combining multi-frequency data. 
We show the architectures in \cref{fig:architecture_baselines}. We also compare our method with the Wide-Band Butterfly Network \citep{li_wide-band_2022,li_accurate_2021}.
For broader context, we refer to \cite{li_wide-band_2022} for comparisons between the Wide-Band Butterfly Network and classical multi-frequency methods that do not involve any machine learning such as Full Waveform Inversion (FWI) and a Least-Squares (LS) scheme. The authors show that the Wide-Band Butterfly Network achieves better accuracy than either FWI or LS with much faster inference times with minimal frequency or hyperparameter tuning.

\newcommand{\ip}[1]{\langle #1 \rangle}

\paragraph{\wideinputnet}
In \wideinputnet, we use an FYNet architecture that is constrained to learn all of the adjoint operators $F_{k_1}^*, \hdots, F_{k_{N_k}}^*$ jointly. 
We concatenate the inputs $[d_{k_1}, \hdots, d_{k_{N_k}}]$ along a new channel dimension, so the input array has shape $(N_m, N_h, N_k, 2)$. The concatenated input is then passed into a standard FYNet architecture, with the first layer having slightly wider convolutional channels. 
The shape of the weights in the 2D convolutional layers is constant in the problem dimensions, so the number of parameters in this network is dominated by the 1D convolutional weights and scales as $O(N_k^2 N_h)$.

\paragraph{\parallelnet}
In \parallelnet, we use an extension of FYNet that allows each adjoint operator $F_{k_1}^*, \hdots, F_{k_{N_k}}^*$ to be learned individually. 
Each input $d_{k_t}$ is input to a unique 1D CNN which emulates $F^*_{k_t}$. 
After the adjoint operators are learned separately, the results are concatenated along a channel dimension, and the filtering operators $\left( F^*_{k_t}F_{k_t} + \mu I \right)^{-1}$ are emulated jointly by 2D CNN layers. 
Again, the number of 2D CNN weights does not scale with the problem dimensions, so the number of parameters in this network is dominated by the $N_k$ 1D CNN blocks and scales as $O(N_k N_h)$. 

\paragraph{Wide-Band Butterfly Network}
The Wide-Band Butterfly Network is introduced and defined in \cite{li_wide-band_2022,li_accurate_2021}. Similar to  \wideinputnet, this architecture also jointly parameterizes the adjoint operators $F_{k_1}^*, \hdots, F_{k_{N_k}}^*$, but it leverages the complementary low-rank property of $F_{k_t}^*$ \citep{khoo_switchnet_2019} to hierarchically merge the data using a butterfly network.
For this network, we use code provided by the authors.\footnote{\url{https://github.com/borongzhang/ISP_baseline}} 
The reference implementation is limited to using data at three incident frequencies, so we only present results in this setting.

\subsection{Stabilizing Reconstruction by Adding Frequencies}
\label{sec:experiments_noiseless}

First, we test whether the intuition built in \cref{sec:our_method} is true in a machine learning context. We test whether machine learning methods that operate on data with multiple incoming wave frequencies are more accurate and stable than single-frequency machine learning methods. 
To make this comparison fair, we create a sequence of training datasets with number of incident wave frequencies $N_k \in \{1, \dots, 5\}$ and keep the amount of training data, $n N_k$, constant for each dataset by suitably adjusting the number of training samples $n$.

For the $N_k = 1$ dataset, we train an FYNet model, and for the other datasets, we train our three models: \wideinputnet , \parallelnet, and \ournet. For each model, we train for a fixed number of epochs and choose the model weights at the epoch at which the error on a validation set of size $n/10$
is minimized. We also use the validation set to search over various hyperparameters, such as the size of 1D and 2D convolutional kernels, the number of channels in the convolutional layers, and optimization hyperparameters, such as step size and weight decay. For the Wide-Band Butterfly Network, we search over the rank of the butterfly factorization, as well as optimization hyperparameters, such as initial learning rate, batch size, and learning rate decay (\cref{sec:appendix_hyperparameters}).

\begin{figure}[!ht]
    \centering
	\includegraphics[width=0.7\linewidth]{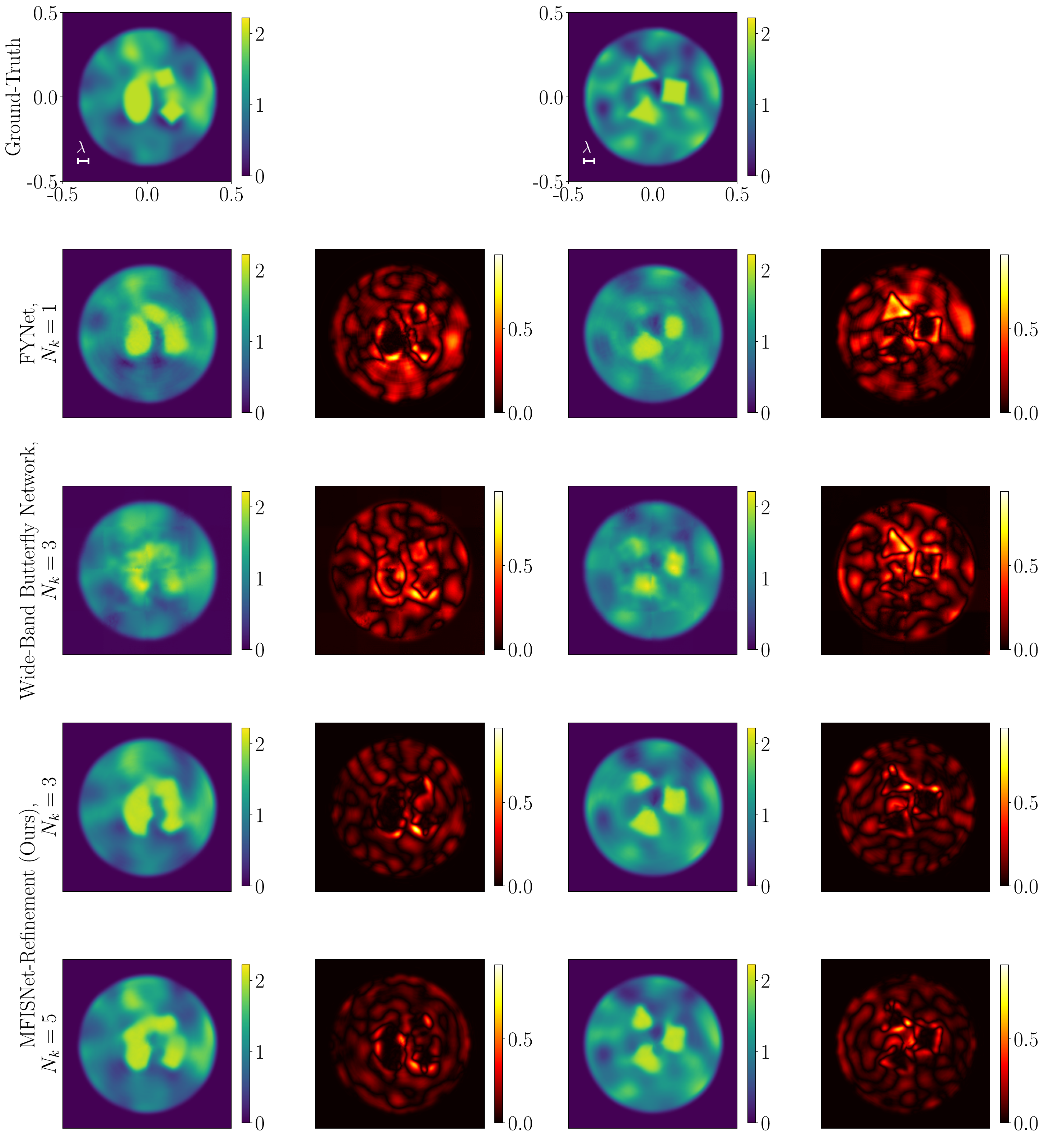}
    \caption{Sample predictions from models trained on different datasets. Predictions and errors on two held-out test samples are shown. The first row shows the ground-truth scattering potential; in this plot we show the wavelength corresponding to the maximum frequency $k=32\pi$. The remaining rows show predictions and errors for FYNet, Wide-Band Butterfly Network, \ournet~$(N_k=3)$, and \ournet~$(N_k =5)$.
    See \cref{sec:appendix_extra_plots} for additional samples and outputs from \parallelnet~and \wideinputnet.}
    \label{fig:target_and_preds}
\end{figure}

\begin{sloppypar}
We present the results of this experiment in \cref{tab:experiment_1} and \cref{fig:target_and_preds,fig:bar_chart}. 
See also \cref{sec:appendix_extra_plots} for additional empirical results, which include training and testing runtimes, visualizations of predictions on more held-out test samples, and visualizations of predictions of \parallelnet\ and \wideinputnet.  
The relatively poor performance of FYNet confirms our belief that we are in a challenging nonlinear problem regime. As more frequencies are added, the multi-frequency methods improve. 
The performances of \wideinputnet\ and \parallelnet\ are comparable, indicating that the distinction between learning adjoint operators separately or jointly does not have a large effect in this setting. 
For $N_k \geq 3$, the tested methods are uniformly outperformed by our method, \ournet. 
As we keep increasing $N_k$, the performance of \ournet~plateaus. 
We hypothesize that our method could be improved by choosing a different set of incident wave frequencies, possibly linearly-spaced in a smaller frequency band. The optimal choice of frequencies could also be learned from data in a reinforcement learning setting \citep{jiang_reinforced_2022}.
\end{sloppypar}

In the case of $N_k=3$, the Wide-Band Butterfly Network underperforms the other methods. 
We hypothesize that the weaker performance of the Wide-Band Butterfly Network may be attributed to several factors: 
on the one hand, the butterfly factorization was inspired by analysis in a weak (linear) scattering regime, but our experiments are in a strong (nonlinear) scattering regime.
Also, the Wide-Band Butterfly Network was previously tested in low-contrast settings with sub-wavelength scatterers and a known background, while we are in an experimental setting with high contrast and an unknown, inhomogeneous background.

\begin{table}[!ht]
    \centering
    \caption*{ \textbf{Performance Comparison (Noiseless)}}
	\resizebox{\linewidth}{!}{
    \begin{tabular}{llllr}
        \toprule
            $N_k$ & ${[k_1, k_2, ...]}$ & $n$ &  \textbf{Method Name} &  \textbf{Relative} $\ell_2$ \textbf{Error}   \\
        \midrule
         1 &  $[32\pi]$ & $10,000$ & FYNet & 
		 $0.159 \pm 0.033$\\
         \midrule
         2 &  $[16\pi, 32\pi]$ & $5,000$ & \wideinputnet & $0.158 \pm 0.030$\\
           & & & \parallelnet & $\mathbf{0.144 \pm 0.029}$\\
           & & & \ournet\ (Ours) & $0.152 \pm 0.032$\\
         \midrule
         3 &  $[8\pi, 16\pi, 32\pi]$ & $3,333$ & Wide-Band Butterfly Network & $0.156 \pm 0.037$\\
          & & & \wideinputnet & $0.121 \pm 0.024$\\
           & & & \parallelnet & $0.107 \pm 0.021$\\
           & & & \ournet\ (Ours) & $\mathbf{0.094 \pm 0.018}$\\
         \midrule
         4 &  $[4\pi, 8\pi, 16\pi, 32\pi]$ & $2,500$ & \wideinputnet & $0.103 \pm 0.020$\\
           & & & \parallelnet & $0.106 \pm 0.021$\\
           & & & \ournet\ (Ours) & $\mathbf{0.082 \pm 0.019}$\\
         \midrule
         5 &  $[2\pi, 4\pi, 8\pi, 16\pi, 32\pi]$ & $2,000$ & \wideinputnet & $0.115 \pm 0.022$\\
           & & & \parallelnet & $0.109 \pm 0.021$\\
           & & & \ournet\ (Ours) & $\mathbf{0.087 \pm 0.018}$\\
         \bottomrule
    \end{tabular}
	}
    \vspace{6pt}
    \caption{When holding the number of forward model evaluations $=nN_k$ constant, methods trained on more frequencies outperform methods with fewer frequencies. The final column reports the relative $\ell_2$ error mean $\pm$ one standard deviation computed over $1,000$ held-out test samples. The lowest mean for each incident frequency set is marked in boldface font.}
    \label{tab:experiment_1}
\end{table}

\begin{figure}[!ht]
    \centering
    \begin{tikzpicture}
        \begin{axis}[
                width  = 0.9 * \linewidth,
                height = 0.65 * 0.9 * \linewidth,
                axis lines*=left,
                ymin = 0,
                ymax = 0.25,
                major x tick style = solid,
                ybar=2*\pgflinewidth,
                bar width=12pt,
                ymajorgrids = true,
                ylabel = {Relative $\ell_2$ error},
                xlabel = {Input frequencies},
                x label style={at={(axis description cs:0.5, -0.4)}, anchor=north},
                symbolic x coords={Baseline,2Freqs,3Freqs,4Freqs,5Freqs},
                xtick = {Baseline, 2Freqs, 3Freqs, 4Freqs, 5Freqs},
                xticklabels = {
                        {$32\pi$},
                        {$16\pi, 32\pi$},
                        {$8\pi, 16\pi, 32\pi$},
                        {$4\pi, 8\pi, 16\pi, 32\pi$},
                        {$2\pi, 4\pi, 8\pi, 16\pi, 32\pi$},
                    },
                ytick = {0, 0.05, 0.1, 0.15, 0.2, 0.25, 0.3},
                yticklabels = {0, 0.05, 0.1, 0.15, 0.2, 0.25, 0.3},
                xticklabel style={rotate=45, font=\small},
                scaled y ticks = false,
                enlarge x limits=0.25,
                legend cell align=left,
                legend style={
                        at={(0.5,0.8)},
                        anchor=south,
                        column sep=1ex
                    },
                legend entries={
                        {FYNet},
                        {Wide-Band Butterfly Net},
                        {\wideinputnet},
                        {\parallelnet},
                        {\ournet\ (\textbf{Ours})},
                    }
            ]
            \addlegendimage{color=bblue, fill=bblue, postaction={pattern=crosshatch dots}};
            \addlegendimage{color=yyellow, fill=yyellow, postaction={pattern=dots}};
            \addlegendimage{color=rred, fill=rred, postaction={pattern=north east lines}};
            \addlegendimage{color=ggreen, fill=ggreen, postaction={pattern={north west lines}}};
            \addlegendimage{color=ppurple, fill=ppurple};
    
            \addplot[
                style={bblue!75!black, fill=bblue, mark=none, postaction={pattern=crosshatch dots}},
                bar shift=0pt,
                error bars/.cd, y dir=both, y explicit, error bar style={color=black},
            ]
            coordinates {(Baseline, 0.159) +- (0, 0.033)};

            \addplot[
                style={rred!75!black, fill=rred, mark=none, postaction={pattern=north east lines}}, bar shift=-12pt,
                error bars/.cd, y dir=both, y explicit, error bar style={color=black},
            ]
            coordinates {
                    (2Freqs, 0.158) +- (0, 0.030)
                    (4Freqs, 0.103) +- (0, 0.020)
                    (5Freqs, 0.115) +- (0, 0.022)
                };
    
            \addplot[
                style={ggreen!75!black, fill=ggreen, mark=none, postaction={pattern=north west lines}}, bar shift=0pt,
                error bars/.cd, y dir=both, y explicit, error bar style={color=black},
            ]
            coordinates {
                    (2Freqs, 0.144) +- (0, 0.029)
                    (4Freqs, 0.106) +- (0, 0.021)
                    (5Freqs, 0.109) +- (0, 0.021)
                };
    
            \addplot[
                style={ppurple!75!black, fill=ppurple, mark=none}, bar shift=12pt,
                error bars/.cd, y dir=both, y explicit, error bar style={color=black},
            ]
            coordinates {
                    (2Freqs, 0.152) +- (0, 0.032)
                    (4Freqs, 0.082) +- (0, 0.019)
                    (5Freqs, 0.087) +- (0, 0.018)
                };
    
            \addplot[
                style={yyellow!75!black, fill=yyellow, mark=none, postaction={pattern=dots}}, bar shift=-18pt,
                error bars/.cd, y dir=both, y explicit, error bar style={color=black},
            ]
            coordinates {
                    (3Freqs, 0.156) +- (0, 0.037)
                };
    
            \addplot[
                style={rred!75!black, fill=rred, mark=none, postaction={pattern=north east lines}}, bar shift=-6pt,
                error bars/.cd, y dir=both, y explicit, error bar style={color=black},
            ]
            coordinates {
                    (3Freqs, 0.121) +- (0, 0.024)
                };
    
            \addplot[
                style={ggreen!75!black, fill=ggreen, mark=none, postaction={pattern=north west lines}}, bar shift=6pt,
                error bars/.cd, y dir=both, y explicit, error bar style={color=black},
            ]
            coordinates {
                    (3Freqs, 0.107) +- (0, 0.021)
                };
    
            \addplot[
                style={ppurple!75!black, fill=ppurple, mark=none}, bar shift=18pt,
                error bars/.cd, y dir=both, y explicit, error bar style={color=black},
            ]
            coordinates {
                    (3Freqs, 0.094) +- (0, 0.018)
                };
    
        \end{axis}
\end{tikzpicture}

    \caption{Illustration of the results of~\cref{tab:experiment_1}. The advantage of the MFISNet-Refinement model grows as more low-frequency data is used, even as the total volume of training data is held constant regardless of the number of frequencies.}
    \label{fig:bar_chart}
\end{figure}

\begin{table}[h]
    \centering
	\resizebox{\linewidth}{!}{
    \begin{tabular}{lllrr}
        \toprule
            $N_k$ &  $n$ &  \textbf{Method Name} &  \textbf{Training Time} & \textbf{Testing Time}   \\
			& & & \textbf{(seconds)} & \textbf{(seconds)} \\
        \midrule
         1  & $10,000$ & FYNet & $870.3$ & $27.3$
		  \\
         \midrule
         3 &   $3,333$ & Wide-Band Butterfly Network & $6,453.6$ & $29.9$ \\
          & & \wideinputnet & $326.2$ & $27.8$ \\
           & & \parallelnet & $256.1$ & $27.8$ \\
           & & \ournet\ (Ours) & $2,033.0$ & $28.2$ \\
         \midrule
         5 & $2,000$ & \wideinputnet & $306.3$ & $27.6$ \\
           & & \parallelnet & $182.2$ & $27.8$ \\
           & & \ournet\ (Ours) & $2,095.5$ & $28.8$ \\
         \bottomrule
    \end{tabular}
	}
    \caption{The time required to train and evaluate machine learning models. In different settings of $N_k$, we report the training time for each method. We also report the time required to make predictions on the entire set of $1,000$ held-out test samples, performed in ten batches of size $100$. As we increase $N_k$, the number of samples $n$ decreases, so \wideinputnet\ and \parallelnet\ see an increase in training speed. For \ournet, as we increase the number of frequencies, we also increase the number of training epochs. 
	The two effects, smaller dataset and more epochs, approximately negate each other, so the training time for \ournet\ remains approximately constant for different $N_k$. 
	The testing times for the FYNet and MFISNet models are all approximately equal because they are dominated by a final polar-to-Cartesian coordinate transformation step, which is performed on a single CPU core but can in principle be accelerated.
    }
    \label{tab:times}
\end{table}

\subsection{Measurement Noise}
\label{sec:experiments_with_noise}
We now turn to the question of robustness against measurement noise. We repeat the experiment above but train and test the models using noisy inputs. We assume an additive noise model similar to \cite{borges_high_2016}. Given a clean input $d_k \in \mathbb{C}^{N_m \times N_h}$ and a desired noise-to-signal ratio $\delta$, we define a noisy input $\tilde{d}_k$ as
\begin{align}
\begin{aligned}[t]
    &\tilde{d}_k = d_k + \sigma (Z_1 + i Z_2), \\
    \text{where} \quad &\sigma = \delta \frac{\| d_k \|_2}{\sqrt{2 N_m  N_h}}; \\
	\text{and} \quad &[Z_j]_{m, h} \overset{\text{i.i.d.}}{\sim} \mathcal{N}(0, 1) \;\; \text{for $j = 1, 2$ and $(m, h) \in [N_m] \times [N_h]$.}
\end{aligned}
\label{eq:additive-noise-model}
\end{align}
Under this noise model, the expected noise-to-signal ratio is $\E_{Z \sim \mathcal{N}(0, I)}[ \| \tilde{d}_k - d_k \| / \| d_k \| ] \approx \delta$.
We do not alter the scattering potentials $q_i$ in any way.
We train and test all models using noisy inputs, and report the results of this experiment in \cref{tab:noise_experiments}. 
For the Wide-Band Butterfly Network, which takes $d_k$ inputs in the original $(r,s)$ coordinates, we add noise to the input in its original coordinates and divide by $\sqrt{2 N_r N_s}$ instead. This results in an equivalent noise profile thanks to the normalization that is grid-invariant.
We note that the models lose between $1\%$ - $2\%$ accuracy on test set, suggesting the methods are relatively robust to the presence of measurement noise. 
Again, our method, \ournet, outperforms all other methods at most values $N_k$. We present the results of this experiment in \cref{tab:noise_experiments} and visualize the results in \cref{fig:bar_chart_noisy}.

\begin{table}[!ht]
    \caption*{\textbf{Performance Comparison (Noisy, $\delta=0.1$)}}
    \centering
	\resizebox{\linewidth}{!}{
    \begin{tabular}{llllr}
        \toprule
            $N_k$ & ${[k_1, k_2, ...]}$ & $n$ &  \textbf{Method Name} &  \textbf{Relative L2 Error}   \\
        \midrule
         1 &  $[32\pi]$ & $10,000$ & FYNet & $0.163 \pm 0.031$\\
         \midrule
         2 &  $[16\pi, 32\pi]$ & $5,000$ & \wideinputnet & $0.164 \pm 0.035$\\
           & & & \parallelnet & $\mathbf{0.148 \pm 0.029}$\\
           & & & \ournet\ (Ours) & $0.151 \pm 0.031$\\
         \midrule
         3 &  $[8\pi, 16\pi, 32\pi]$ & $3,333$ & Wide-Band Butterfly Network & $0.156 \pm 0.037$\\
           & & & \wideinputnet & $0.125 \pm 0.025$\\
           & & & \parallelnet & $ 0.110 \pm 0.023$\\
           & & & \ournet\ (Ours) & $\mathbf{0.097 \pm 0.019}$\\
         \midrule
         4 &  $[4\pi, 8\pi, 16\pi, 32\pi]$ & $2,500$ & \wideinputnet & $0.102 \pm 0.021$\\
           & & & \parallelnet & $0.108 \pm 0.019$\\
           & & & \ournet\ (Ours) & $\mathbf{0.086 \pm 0.019}$\\
         \midrule
         5 &  $[2\pi, 4\pi, 8\pi, 16\pi, 32\pi]$ & $2,000$ & \wideinputnet & $0.112 \pm 0.022$\\
           & & & \parallelnet & $0.103 \pm 0.020$\\
           & & & \ournet\ (Ours) & $\mathbf{0.090 \pm 0.017}$\\
         \bottomrule
    \end{tabular}
	}
    \vspace{6pt}
    \caption{Model evaluation on noisy train and test inputs with $\delta=0.1$ (see \eqref{eq:additive-noise-model} for a description of the noise model). The final column reports the relative $\ell_2$ error mean $\pm$ one standard deviation computed over $1,000$ held-out test samples. The lowest mean for each incident frequency set is marked in boldface font.}
    \label{tab:noise_experiments}
\end{table}

\begin{figure}[!ht]
    \centering
    \begin{tikzpicture}
        \begin{axis}[
            width  = 0.9 * \textwidth,
            height = 0.65 * 0.9 * \textwidth,
            axis lines*=left,
            ymin = 0,
            ymax = 0.25,
            major x tick style = solid,
            ybar=2*\pgflinewidth,
            bar width=12pt,
            ymajorgrids = true,
            ylabel = {Relative $\ell_2$ error},
            xlabel = {Input frequencies},
            x label style={at={(axis description cs:0.5, -0.4)}, anchor=north},
            symbolic x coords={Baseline,2Freqs,3Freqs,4Freqs,5Freqs},
            xtick = {Baseline, 2Freqs, 3Freqs, 4Freqs, 5Freqs},
            xticklabels = {
                    {$32\pi$},
                    {$16\pi, 32\pi$},
                    {$8\pi, 16\pi, 32\pi$},
                    {$4\pi, 8\pi, 16\pi, 32\pi$},
                    {$2\pi, 4\pi, 8\pi, 16\pi, 32\pi$},
                },
            ytick = {0, 0.05, 0.1, 0.15, 0.2, 0.25, 0.3},
            yticklabels = {0, 0.05, 0.1, 0.15, 0.2, 0.25, 0.3},
            xticklabel style={rotate=45, font=\small},
            scaled y ticks = false,
            enlarge x limits=0.25,
            legend cell align=left,
            legend style={
                    at={(0.5,0.8)},
                    anchor=south,
                    column sep=1ex
                },
            legend entries={
                    {FYNet},
                    {Wide-Band Butterfly Net},
                    {\wideinputnet},
                    {\parallelnet},
                    {\ournet\ (\textbf{Ours})},
                }
        ]
        \addlegendimage{color=bblue, fill=bblue, postaction={pattern=crosshatch dots}};
        \addlegendimage{color=yyellow, fill=yyellow, postaction={pattern=dots}};
        \addlegendimage{color=rred, fill=rred, postaction={pattern=north east lines}};
        \addlegendimage{color=ggreen, fill=ggreen, postaction={pattern={north west lines}}};
        \addlegendimage{color=ppurple, fill=ppurple};
    
        \addplot[
            style={bblue!75!black, fill=bblue, mark=none, postaction={pattern=crosshatch dots}},
            bar shift=0pt,
            error bars/.cd, y dir=both, y explicit, error bar style={color=black},
        ]
        coordinates {(Baseline, 0.163) +- (0, 0.031)};

        \addplot[
            style={rred!75!black, fill=rred, mark=none, postaction={pattern=north east lines}}, bar shift=-12pt,
            error bars/.cd, y dir=both, y explicit, error bar style={color=black},
        ]
        coordinates {
                (2Freqs, 0.164) +- (0, 0.035)
                (4Freqs, 0.102) +- (0, 0.021)
                (5Freqs, 0.112) +- (0, 0.022)
            };
    
        \addplot[
            style={ggreen!75!black, fill=ggreen, mark=none, postaction={pattern=north west lines}}, bar shift=0pt,
            error bars/.cd, y dir=both, y explicit, error bar style={color=black},
        ]
        coordinates {
                (2Freqs, 0.148) +- (0, 0.029)
                (4Freqs, 0.108) +- (0, 0.019)
                (5Freqs, 0.103) +- (0, 0.020)
            };
    
        \addplot[
            style={ppurple!75!black, fill=ppurple, mark=none}, bar shift=12pt,
            error bars/.cd, y dir=both, y explicit, error bar style={color=black},
        ]
        coordinates {
                (2Freqs, 0.151) +- (0, 0.031)
                (4Freqs, 0.086) +- (0, 0.019)
                (5Freqs, 0.090) +- (0, 0.017)
            };
    
        \addplot[
            style={yyellow!75!black, fill=yyellow, mark=none, postaction={pattern=dots}}, bar shift=-18pt,
            error bars/.cd, y dir=both, y explicit, error bar style={color=black},
        ]
        coordinates {
                (3Freqs, 0.156) +- (0, 0.037)
            };
    
        \addplot[
            style={rred!75!black, fill=rred, mark=none, postaction={pattern=north east lines}}, bar shift=-6pt,
            error bars/.cd, y dir=both, y explicit, error bar style={color=black},
        ]
        coordinates {
                (3Freqs, 0.125) +- (0, 0.025)
            };
    
        \addplot[
            style={ggreen!75!black, fill=ggreen, mark=none, postaction={pattern=north west lines}}, bar shift=6pt,
            error bars/.cd, y dir=both, y explicit, error bar style={color=black},
        ]
        coordinates {
                (3Freqs, 0.110) +- (0, 0.023)
            };
    
        \addplot[
            style={ppurple!75!black, fill=ppurple, mark=none}, bar shift=18pt,
            error bars/.cd, y dir=both, y explicit, error bar style={color=black},
        ]
        coordinates {
                (3Freqs, 0.097) +- (0, 0.019)
            };
    
    \end{axis}
    
    \end{tikzpicture}
    \caption{Illustration of results in the noisy measurement regime (\cref{tab:noise_experiments}). The relative performance of the models remains the same as in the noiseless case.}
    \label{fig:bar_chart_noisy}
\end{figure}

\subsection{Investigating the Training Method}
\label{sec:experiments_ablation_study}
\begin{sloppypar}
Our method unfolds the reconstruction problem into a sequence of simpler frequency-dependent refinement steps. This emulates the sequential, frequency-dependent structure of the recursive linearization method. Compared to existing methods, our method introduces both a new residual architecture and a new sequential training procedure, so a natural question would be how important each of these factors is to our method's success.
To test this question, we investigate two different changes to our training method which remove the progressive refinement structure. We describe each adjustment below. The results are presented in Table \ref{tab:ablation study}. 
In this experiment, we find the accuracy of our method is relatively robust to perturbations of the training method presented in \cref{alg:training_procedure}. This indicates the advantage of our approach is primarily driven by the residual architecture. 

\end{sloppypar}

\paragraph{No Homotopy through Frequency} Rather than sequentially training each network block after the previous blocks have been optimized, we jointly train all of the blocks by optimizing the loss function 
    \begin{align}
        \| \hat{q}_{k_{N_{k}}} -  q \|_2^2 + \sum_{t=1}^{N_k - 1} \gamma^{N_k - t} \| \hat{q}_{k_t} - \lpf_{2k_t} q \|_2^2 
        \label{eq:no_sequential_pretraining}
    \end{align}
    Here $\gamma$ is a hyperparameter which controls the relative importance of the different loss terms. We tuned over a few choices of $\gamma$; see \cref{tab:no_homotopy_hyperparams} for details. 
\paragraph{No Progressive Refinement} Instead of using the intermediate loss terms  $\| \hat{q}_{k_t} - \lpf_{2k_t} q \|_2^2$, designed to promote specific network blocks learning different parts of the reconstruction, we train the network by only optimizing the final loss term $\| \hat{q}_{k_{N_{k}}} - q \|_2^2$. This is the standard training loss used in most other works, including \cite{khoo_switchnet_2019,fan_solving_2019,li_wide-band_2022}.
\begin{table}[h]
    \centering
	\resizebox{\linewidth}{!}{
    \begin{tabular}{lrrr}
        \toprule
            \textbf{Training Method}  & ${[k_1, k_2, ...]}$ & $n$ & \textbf{Relative L2 Error}\\
        \midrule

        No Progressive Refinement &  $[ 16\pi, 32\pi]$ & $5,000$ & $\mathbf{0.152\pm 0.030}$\\
        No Homotopy through Frequency &  & & $\mathbf{0.152\pm 0.030}$\\
        \cref{alg:training_procedure} &  & & $\mathbf{0.152\pm 0.032}$\\
		\midrule
        No Progressive Refinement &  $[ 8 \pi, 16\pi, 32\pi]$ & $3,333$ & $\mathbf{0.094 \pm 0.019}$\\
        No Homotopy through Frequency &  & & $0.097 \pm 0.018$\\
        \cref{alg:training_procedure} &  & & $\mathbf{0.094 \pm 0.018}$\\ 
        \midrule
        No Progressive Refinement &  $[4 \pi, 8 \pi, 16\pi, 32\pi]$ & $2,500$ & $0.091 \pm 0.020$\\
        No Homotopy through Frequency &  & & $0.086 \pm 0.018$\\
        \cref{alg:training_procedure} &  & & $\mathbf{0.082 \pm 0.019}$\\
        \midrule
        No Progressive Refinement & $[ 2\pi, 4\pi, 8\pi, 16\pi, 32\pi]$ & $2,000$ & $\mathbf{0.082 \pm 0.018}$\\
        No Homotopy through Frequency & & & $0.087 \pm 0.016$\\
        \cref{alg:training_procedure} & & & $0.087 \pm 0.018$\\

        \bottomrule

    \end{tabular}
	}
    \caption{Alternative training methods, designed to remove parts of the recursive linearization structure, produce models of similar accuracy to the training procedure discussed in \cref{alg:training_procedure}. This indicates the advantage of our \ournet\ method is driven primarily by its residual architecture. We describe the alternate training methods in \cref{sec:experiments_ablation_study}. The optimal hyperparameters for these models are listed in \cref{sec:appendix_hyperparameters}.}
    \label{tab:ablation study}
\end{table}

\subsection{Investigating the Speed of Training Convergence}
\label{sec:convergence_speed}
\begin{sloppypar}
While \cref{tab:experiment_1,tab:noise_experiments} show that \ournet\ is more accurate than other networks in most settings, \cref{tab:times} shows that training our network is slower than the baseline \parallelnet\ and \wideinputnet\ architectures. 
This is because our \ournet\ architecture is more complex than \parallelnet\ or \wideinputnet, and our training procedure (\cref{alg:training_procedure}) takes a block-wise approach which requires many training epochs. 
In this section, we conduct experiments to investigate whether we can reduce the time spent training \ournet. 
Because the results presented in \cref{sec:experiments_ablation_study} suggest that all three training methods  considered, \cref{alg:training_procedure}, ``No Homotopy through Frequency'', and ``No Progressive Refinement'', produce models which are approximately similar in accuracy, we consider training with all of these methods.
\end{sloppypar}

\begin{sloppypar}
We introduce another training method, designed to mimic \cref{alg:training_procedure} but converge in fewer training epochs. 
This method, which we call ``\cref{alg:training_procedure} + Warm-Start Initialization'', uses the previous block's parameters $\theta_{t-1}$ to initialize the weights of the next block $\theta_{t}$. 
This ``warm start'' is inspired by the fact that neighboring blocks in our \ournet\ architecture are designed to learn similar functions; each block is meant to increase the resolution of the estimated scattering potential by a small amount. 
Thus, the weights learned by the previous block may be a good starting point for the optimization of the next block. 
We give full pseudocode for this algorithm in \cref{sec:appendix_algo}.
\end{sloppypar}

\begin{sloppypar}
We set a early stopping criterion, which terminates training if the relative $\ell_2$ error has not decreased by at least $10^{-3}$ over the past 15 epochs. We use this early stopping condition to terminate the ``No Homotopy through Frequency'' and ``No Progressive Refinement'' training methods. Similarly, we use this early stopping criterion to terminate each block in \cref{alg:training_procedure}. In \cref{tab:training_convergence_speed}, we measure the number of epochs used and runtime of each training method. 
We observe that the warm-start initialization decreases the number of epochs and training time of \cref{alg:training_procedure}. For $N_k=3$ frequencies, this method converges faster than the other three methods. 
In both cases, using the warm start initialization incurs a slight decrease in accuracy while greatly increasing the speed of training convergence. Different early stopping criteria could achieve different tradeoffs between training speed and accuracy, and we leave the investigation of such criteria to future work. 
\end{sloppypar}

\begin{sloppypar}
The time required to complete an epoch of training for \cref{alg:training_procedure} with and without warm-starting is much shorter than that of the ``No Progressive Refinement'' and ``No Homotopy through Frequency'' methods. 
This is because for the majority of epochs in \cref{alg:training_procedure}, only one block is updated at a time, making the backpropogation of gradients much faster. In the ``No Progressive Refinement'' and ``No Homotopy through Frequency'' methods, all trainable parameters are updated in each epoch.
\end{sloppypar}

\begin{table}[h]
    \centering
	\resizebox{\linewidth}{!}{
    \begin{tabular}{lrrrrr}
        \toprule
            \textbf{Training Method}  & ${[k_1, k_2, ...]}$ & $n$ & \textbf{Number of} & \textbf{Training Time} &  \textbf{Relative} $\ell_2$ \\
			& & & \textbf{Epochs} & \textbf{(seconds)} &  \textbf{Error} \\
        \midrule
        No Progressive Refinement &  $[ 8 \pi, 16\pi, 32\pi]$ & $3,333$ & $90$ & $1,200.6$ & $0.096 \pm 0.020$ \\
        No Homotopy through Frequency & & & $55$  & $920.3$ & $0.105 \pm 0.020$  \\
        \cref{alg:training_procedure} &  & & $215$ & $1,226.4$ & $0.094 \pm 0.019$  \\ 
        \cref{alg:training_procedure} + Warm-Start Init. &  & & $195$  & $802.8$ & $0.098 \pm 0.019$  \\ 
        \midrule
        No Progressive Refinement & $[ 2\pi, 4\pi, 8\pi, 16\pi, 32\pi]$ & $2,000$ & $45$ & $811.6$ & $0.095 \pm 0.019$ \\
        No Homotopy through Frequency & & & $50$ & $997.8$ & $0.091 \pm 0.017$ \\
        \cref{alg:training_procedure} & & & $260$ & $1,417.5$ & $0.087 \pm 0.017$ \\
        \cref{alg:training_procedure} + Warm-Start Init. & & & $210$ & $1041.9$ & $0.105 \pm 0.020$ \\
        \bottomrule
    \end{tabular}
	}
    \caption{Using warm-starting in \cref{alg:training_procedure} reduces the training runtime between $20$ and $35\%$. In this table, we report time and epochs required to train \ournet~ models with different training methods, and the relative $\ell_2$ errors of each fully-trained model. The warm-start initialization method and the convergence criterion are described in \cref{sec:convergence_speed}.}
    \label{tab:training_convergence_speed}
\end{table}

%% file: revised_conclusion.tex
\section{Conclusion}
\label{sec:conclusion}

This paper investigates the use of multi-frequency data in deep learning approaches to the inverse medium scattering problem in a highly nonlinear, full-aperture regime. 
We review standard optimization results for this problem, identify recursive linearization as an algorithm particularly well-suited for this problem, and use this insight to design a neural network architecture and training method.
We experimentally evaluate our proposed approach, comparing against novel and previously-published methods for combining multi-frequency data.
In these comparisons, we find our method outperforms the other methods across a wide range of data settings, including with and without measurement noise.

This work also leaves open important questions about machine learning in different multi-frequency data settings. 
We leave the investigation of these important problems to future work.
The first setting is seismic imaging, where sources and receivers are located on one side of the scattering potential, resulting in limited-aperture measurements.
Another setting to consider is full-aperture measurements, with a small fixed or frequency-dependent number of source and receiver directions, as 
$N_s$ and $N_r$ drive real-world costs when implementing an imaging system.
Real-world imaging systems often encounter noise that is not well-approximated by an additive zero-mean Gaussian noise model; characterizing the robustness of deep learning methods in more realistic noise settings is important future work.
Finally, we suggest a distribution of scattering potentials with an unknown, smoothly varying background occluded by strongly scattering shapes. We note that an important open problem is to learn to \emph{segment} the reconstruction into disjoint regions, containing only background or only the strong scatterers. 

%% file: revised_contribution.tex
\section*{Author Contributions}
\begin{itemize}
\item 
\textbf{Owen Melia:} Conceptualization; Data curation; Methodology; Software; Writing - original draft. 
\item \textbf{Olivia Tsang:} Conceptualization; Data curation; Methodology; Software; Writing - original draft. 
\item \textbf{Vasileios Charisopoulos:} Conceptualization; Methodology; Supervision; Writing - review \& editing. 
\item \textbf{Yuehaw Khoo:} Conceptualization; Methodology; Supervision; Writing - review \& editing. 
\item \textbf{Jeremy Hoskins:} Conceptualization; Methodology; Supervision; Writing - review \& editing. 
\item \textbf{Rebecca Willett:} Conceptualization; Methodology; Supervision; Funding acquisition; Project administration; Writing - review \& editing. 
\end{itemize}

%% file: revised_data_availability.tex
\section*{Data Availability}
Our publicly-available GitHub repository \url{https://github.com/meliao/mfisnets} contains the following:
\begin{itemize}
    \item Code for defining and training \ournet, \parallelnet, and \wideinputnet.
    \item Code for generating the dataset used in our experiments.
\end{itemize}
Our publicly-available dataset of $10,000$ training samples, $1,000$ validation samples, and $1,000$ test samples can be downloaded from \url{https://doi.org/10.5281/zenodo.14514353}. 

%% file: revised_acknowledgement.tex
\section*{Acknowledgements}
The authors would like to thank Borong Zhang for sharing his implementation of the Wide-Band Butterfly Network.
OM, OT, VC, and RW gratefully acknowledge the support of AFOSR FA9550-18-1-0166 and NSF DMS-2023109. RW and YK gratefully acknowledge the support of DOE DE-SC0022232. YK gratefully acknowledges the support of NSF DMS-2339439. The team gratefully acknowledges the support of the Margot and Tom Pritzker Foundation.

%% file: revised_appendix_data_generation.tex
\section{Distribution of Scattering Potentials}
\label{sec:appendix_data_gen}
To generate samples from $\mathcal{D}$, our distribution of scattering potentials, we draw a random smoothly-varying background and three shapes with random sizes, positions, and rotations. 
This section provides details about the generation of these scattering objects.

The random low-frequency backgrounds were generated by drawing random Fourier coefficients and filtering out the high frequencies using %
$\lpf_{7.\bar{1}\pi}$.
The resulting background was transformed to Cartesian coordinates, and then shifted and scaled so the maximum value was $2.0$ and the minimum value was $0.0$. The background was truncated to $0.0$ outside of the disk of radius $0.4$.
Three shapes were randomly chosen among equilaterial triangles, squares, and ellipses. The three shapes had randomly-chosen centers and rotations, constrained to be non-overlapping and fit inside the disk of radius $0.4$. The side lengths of the squares and triangles were uniformly sampled from $[0.1, 0.15]$. The major axis lengths of the ellipses were uniformly sampled from $[0.1, 0.15]$, and the minor axis lengths were uniformly sampled from $[0.05, 0.1]$. Finally, $\lpf_{32\pi}$ was applied to the scattering potential.

%% file: revised_appendix_hyperparameter_optimization.tex
\section{Hyperparameter Search}
\label{sec:appendix_hyperparameters}
For our hyperparameter searches, we trained models on a grid of hyperparameters and evaluated them on a validation set every $5$ epochs. 
We found the epoch and hyperparameter setting which produced the lowest error on the validation set, and used those model weights for final evaluation on a held-out test set. 
\subsection{FYNet and MFISNet Models}
We train all models using the Adam algorithm, with a batch size of $16$ samples. 
All of the FYNet, \parallelnet, \wideinputnet, and \ournet\ models tested have 3 1D convolutional layers followed by 3 2D convolutional layers; ReLU activations are used between layers.
In the FYNet, \parallelnet\ and \ournet\ models, we use 1D and 2D convolutional kernels with $24$ channels following \citep{fan_solving_2019}; in the \wideinputnet\ models, we use 1D and 2D convolutional kernels with $24N_k$ channels to adjust for the increased input size.
To train \wideinputnet, \parallelnet\ and \ournet, we search over a grid of architecture and optimization hyperparameters. We report the optimal hyperparameters we found in \cref{tab:fynet_hyperparams,tab:wide_input_fynet_hyperparams,tab:parallel_fynet_hyperparams,tab:residual_fynet_hyperparams,tab:no_homotopy_hyperparams}. %

\begin{sloppypar}
The FYNet model and all MFISNet models use no input or output normalization. 
The parameters for the 1D convolutional layers were initialized by drawing from a uniform distribution $U[0, 2/d_{in}]$, where $d_{in}$ is the input dimension for a given layer. 
The parameters for the 2D convolutional layers were initialized by the standard initialiaztion scheme for  2D convolutional layers in \texttt{Pytorch}, 
We performed preliminary investigation into different initialization schemes and found that the initialization scheme had a smaller effect than architecture or optimization hyperparameters. 
As a result, we decided to restrict our hyperparameter search to the architecture and optimization hyperparameters.

\end{sloppypar}

\begin{description}
    \item[1d kernel size] This is the number of frequency components in the 1D convolutional filters emulating $F_k^*$. We search over values $\set{20, 40, 60}$.
    \item[2d kernel size] This is the size (in pixels) of the 2D convolutional kernel used in the layers emulating $\left(F_k^* F_k + \mu I\right)^{-1}$. We search over values $\set{5, 7}$.
    \item[Weight decay] The weight decay parameter adds an $\ell_2$ weight regularization term to the loss function. This hyperparameter determines the coefficient of this regularization term. We search over values $\set{0.0, 1\times 10^{-3}}$.
    \item[Learning rate] This is the step size for the Adam optimization algorithm. We search over values $\set{1\times 10^{-4}, 5 \times 10^{-4}, 1 \times 10^{-3}}$. 
    \item[LR decrease] For \ournet\ models only, we decrease the learning rate each time we begin training a new network block. This parameter determines the multiplicative decrease that we apply to the learning rate. We search over values $\set{1.0, 0.25}$.
\end{description}

\begin{table}[!ht]
    \centering
    \begin{tabular}{l|l||l}
        \toprule
        \textbf{Hyperparameter} & \multicolumn{2}{c}{\textbf{Data Setting (FYNet)}} \\
         & $\delta=0.0; N_k = 1$ & $\delta=0.1; N_k = 1$   \\
        \midrule
        \textbf{1d kernel size} & 60 & 60 \\
        \textbf{2d kernel size} & 5 & 5 \\
        \textbf{Weight decay} & $0.0$ & $0.0$ \\
        \textbf{Learning Rate} & $1\times 10^{-3}$ & $1\times 10^{-3}$ \\
        \bottomrule
    \end{tabular}
    \caption{Optimal hyperparameters for FYNet.}
    \label{tab:fynet_hyperparams}
\end{table}

\begin{table}[!ht]
    \centering
    \resizebox{\textwidth}{!}{
    \begin{tabular}{l|l|l|l|l||l|l|l|l}
        \toprule
        \textbf{Hyperparameter} & \multicolumn{8}{c}{\textbf{Data Setting (\ournet)}} \\
           & $N_k = 2$ & $N_k = 3$ & $ N_k =4$ & $ N_k =5$ & $N_k = 2$ & $N_k = 3$ & $ N_k =4$ & $ N_k =5$   \\
           & \multicolumn{4}{c}{$\delta = 0.0$}  & \multicolumn{4}{c}{$\delta = 0.1$}\\
        \midrule
           
           \textbf{1d kernel size} & $40$ & $20$ & $40$ & $40$ & $40$ & $20$ & $40$ & $20$ \\
           \textbf{2d kernel size} & $7$ & $5$ & $5$ & $5$  & $5$ & $5$  & $5$ & $5$ \\
           \textbf{Weight decay} & $1\times10^{-3}$ & $0.0$ & $1\times 10^{-3}$ & $0.0$  & $0.0$  & $0.0$ & $1\times10^{-3}$ & $1 \times 10^{-3}$  \\
           \textbf{Learning rate} & $5 \times 10^{-4}$ & $5 \times 10^{-4}$ & $1 \times 10^{-3}$ & $5\times10^{-4}$ & $1\times 10^{-3}$ & $5\times10^{-4}$ & $1 \times 10^{-4}$ & $1 \times 10^{-4}$ \\
           \textbf{LR decrease} & $0.25$ & $1.0$ & $0.25$ & $1.0$ & $0.25$ & $1.0$ & $1.0$ & $1.0$  \\
        \bottomrule
    \end{tabular}
    }
    \caption{Optimal hyperparameters for \ournet.}
    \label{tab:residual_fynet_hyperparams}
\end{table}

\begin{table}[!ht]
    \centering
    \resizebox{\textwidth}{!}{
    \begin{tabular}{l|l|l|l|l||l|l|l|l}
        \toprule
        \textbf{Hyperparameter} & \multicolumn{8}{c}{\textbf{Data Setting (\parallelnet)}} \\
           & $N_k = 2$ & $N_k = 3$ & $ N_k =4$ & $ N_k =5$ & $N_k = 2$ & $N_k = 3$ & $ N_k =4$ & $ N_k =5$   \\
           & \multicolumn{4}{c}{$\delta = 0.0$}  & \multicolumn{4}{c}{$\delta = 0.1$}\\
        \midrule
           
           \textbf{1d kernel size} & $40$ & $20$ & $20$ & $20$ & $40$ & $40$ & $20$ & $20$ \\
           \textbf{2d kernel size} & $5$ & $7$ & $5$ & $7$ & $5$ & $5$ & $7$ & $7$ \\
           \textbf{Weight decay} & $0.0$  & $1 \times 10^{-3}$ & $0.0$ & $1 \times 10^{-3}$ & $1 \times 10^{-3}$ & $1\times 10^{-3}$ & $1 \times 10^{-3}$ & $0.0$ \\
           \textbf{Learning rate} & $5 \times 10^{-4}$  & $5 \times 10^{-4}$ & $1 \times 10^{-3}$ & $1 \times 10^{-3}$  & $1 \times 10^{-3}$ & $1 \times 10^{-3}$ & $1 \times 10^{-3}$ & $5 \times 10^{-4}$ \\
        \bottomrule
    \end{tabular}
    }
    \caption{Optimal hyperparameters for \parallelnet.}
    \label{tab:parallel_fynet_hyperparams}
\end{table}

\begin{table}[!ht]
    \centering
    \resizebox{\textwidth}{!}{
    \begin{tabular}{l|l|l|l|l||l|l|l|l}
        \toprule
        \textbf{Hyperparameter} & \multicolumn{8}{c}{\textbf{Data Setting (\wideinputnet)}} \\
           & $N_k = 2$ & $N_k = 3$ & $ N_k =4$ & $ N_k =5$ & $N_k = 2$ & $N_k = 3$ & $ N_k =4$ & $ N_k =5$   \\
           & \multicolumn{4}{c}{$\delta = 0.0$}  & \multicolumn{4}{c}{$\delta = 0.1$}\\
        \midrule
           \textbf{1d kernel size} & 40 & 20 & 20 & 40 & 20 & 20 & 20 & 40 \\
           \textbf{2d kernel size} & 5 & 5 & 5 & 5 & 7 & 5 & 5 & 5 \\
           \textbf{Weight decay} & 0 & $1\times 10^{-3}$ & 0 & 0 & $1\times 10^{-3}$ & $1\times 10^{-3}$ & $1\times 10^{-3}$ & $1\times 10^{-3}$ \\
           \textbf{Learning Rate} & $5\times 10^{-4}$ & $5\times 10^{-4}$ & $1\times 10^{-4}$ & $1\times 10^{-3}$ & $1\times 10^{-3}$ & $5\times 10^{-4}$ & $1\times 10^{-4}$ & $1\times 10^{-3}$ \\
        \bottomrule
    \end{tabular}
    }
    \caption{Optimal hyperparameters for \wideinputnet.}
    \label{tab:wide_input_fynet_hyperparams}
\end{table}

\begin{table}[!ht]
    \centering
    \begin{tabular}{l|l|l|l|l}
        \toprule
        \textbf{Hyperparameter} & \multicolumn{4}{c}{\textbf{Data Setting (\ournet)}} \\
        &  \multicolumn{4}{c}{\textbf{with No Progressive Refinement Training}} \\
        & $N_k=2$  & $N_k=3$ & $N_k=4$ & $N_k=5$   \\
        \midrule
           
           \textbf{1d kernel size} & $40$ & $20$ & $40$ & $40$   \\
           \textbf{2d kernel size} & $5$ & $5$ & $5$ & $7$   \\
           \textbf{Weight decay} & $0.0$ & $1\times 10^{-3}$ & $1\times10^{-3}$ & $0.0$    \\
           \textbf{Learning rate} & $5\times 10 ^{-4}$ & $5 \times 10^{-4}$ & $1 \times 10^{-4}$ & $1 \times 10^{-4}$  \\
        \bottomrule
    \end{tabular}
    \caption{Optimal hyperparameters for the \ournet\ models trained with the ``No Progressive Refinement'' training condition (\cref{sec:experiments_ablation_study}). All of the models were trained on noiseless training samples.}
    \label{tab:no_progressive_refinement_hyperparams}
\end{table}

\begin{table}[!ht]
    \centering
    \resizebox{\textwidth}{!}{
    \begin{tabular}{l|l|l|l|l}
        \toprule
        \textbf{Hyperparameter} & \multicolumn{4}{c}{\textbf{Data Setting (\ournet)}} \\
        &  \multicolumn{4}{c}{\textbf{with No Homotopy through Frequency Training}} \\
        & $N_k=2$  & $N_k=3$ & $N_k=4$ & $N_k=5$   \\
        \midrule
           
           \textbf{1d kernel size} &  $40$  & $20$ & $40$ & $40$ \\
           \textbf{2d kernel size} &  $5$ & $7$ & $7$ & $7$ \\
           \textbf{Weight decay} & $1\times 10^{-3}$  & $1\times 10^{-3}$ & $1\times 10^{-3}$ & $1 \times 10^{-3}$ \\
           \textbf{Learning rate} & $5\times 10^{-4}$ & $5\times 10^{-4}$ & $5\times 10^{-4}$ & $1\times 10^{-4}$ \\
           $\mathbf{\gamma}$ & $1.0$ & $1.0$ & $1.1$ & $1.1$ \\
        \bottomrule
    \end{tabular}
    }
    \caption{Optimal hyperparameters for the \ournet\ models trained with the ``No Homotopy through Frequency'' training condition (\cref{sec:experiments_ablation_study}). Here, $\mathbf{\gamma}$ is the factor which weights different loss terms (cf.~\eqref{eq:no_sequential_pretraining}). We searched over values $\gamma = \set{0.9, 1.0, 1.1}$. All of the models were trained on noiseless training samples.}
    \label{tab:no_homotopy_hyperparams}
\end{table}

\subsection{Wide-Band Butterfly Network}
To find the optimal Wide-Band Butterfly Network, we optimized over the following hyperparameters. We defined the grid of hyperparameters by taking the original hyperparameter from \cite{li_wide-band_2022} and adding both higher and lower values, where possible. See \cref{tab:widebnet_hyperparams} for the selected values.
\begin{description}
    \item[Rank] This parameter controls the rank of the compression of local patches in the butterfly factorization. Increasing this rank parameter increases the number of learnable parameters in the part of the network emulating $F_k^*$. We found that increasing the rank decreased the train and validation errors, and we increased the rank until we were unable to fit the model and data onto a single GPU. We searched over values $\set{2, 3, 5, 10, 15, 20, 30, 50}$.
    \item[Initial Learning Rate] We decrease the learning rate by a multiplicative factor after $2,000$ minibatches, as suggested by~\cite{li_wide-band_2022}. This parameter is the initial learning rate for the optimization algorithm. We searched over values $\set{5 \times 10^{-4}, 1 \times 10^{-3}, 5 \times 10^{-3}}$.
    \item[Learning Rate Decay] This is the multiplicative decay parameter for the learning rate schedule. We searched over values $\set{0.85, 0.95}$.
    \item[Sigma] \cite{li_wide-band_2022} suggest training the network to match slightly-filtered versions of the ground-truth $q$. This is performed by applying a Gaussian filter to the targets $q^{(i)}$ before training. Sigma is the standard deviation of this Gaussian filter. We do not blur the targets in the test set. We searched over values $\set{0.75, 1.125, 1.5}$.
    \item[Batch Size] This is the number of samples per minibatch. We searched over values $\set{16, 32}$.
\end{description}

\begin{table}[!ht]
    \centering
    \begin{tabular}{l|l|l}
        \toprule
        \textbf{Hyperparameter} & \multicolumn{2}{c}{\textbf{Data Setting}} \\
           & $\delta=0.0; N_k = 3$ & $\delta=0.1; N_k = 3$   \\
        \midrule
        \textbf{Rank} & $50$ & $50$ \\
        \textbf{Initial Learning Rate} & $5\times10^{-4}$ & $5\times10^{-4}$ \\
        \textbf{Learning Rate Decay} & $0.85$ & $0.95$ \\
        \textbf{Sigma} & $1.5$ & $1.5$ \\
        \textbf{Batch Size} & $16$ & $32$ \\
         \bottomrule
    \end{tabular}
    \caption{Optimal hyperparameters for Wide-Band Butterfly Networks.}
    \label{tab:widebnet_hyperparams}
\end{table}

%% file: appendix_warm_start_algo.tex
\section{Training Method with Warm-Start Initialization}
\label{sec:appendix_algo}
\begin{sloppypar}
In this section, we give details of the ``\cref{alg:training_procedure} + Warm-Start Initialization'' training procedure described in \cref{sec:convergence_speed}. 
This training procedure leverages the intuition that sequential blocks in our \ournet\ architecture learn similar functions. In \cref{alg:warm_start_training} we  give full pseudocode for the warm-start training procedure. In this pseudocode, we refer to the trainable parameters in block $t$ as $\theta_t$. If $t>1$, these parameters contain the parameters for the FYNet block, which we call $\theta_{t,1}$ and the parameters for the extra filtering layers, which we call $\theta_{t,2}$.
\end{sloppypar}

\begin{algorithm}[h!]
    \DontPrintSemicolon
        \KwIn{Training data samples $\mathcal{D}_{n} := \left\{
                (q^{(j)}, d_{k_1}^{(j)}, \hdots, d_{k_{N_{k}}}^{(j)})
            \right\}_{j=1}^n$.}
        \For{$t=1,...,N_{k}$}{
            \uIf{$t=1$}{
                Initialize $t_{1,1}$ randomly
            }
            \uElseIf{$t=2$}{
                Initialize $\theta_{2,1}$ with $\theta_{1,1}$ \\
                Initialize $\theta_{2,2}$ randomly
            }
            \Else{
                Initialize $\theta_{t,1}$ with $\theta_{t-1, 1}$ \\
                Initialize $\theta_{t,2}$ with $\theta_{t-1, 2}$ \\
            }
            Set $\theta_t$ as trainable, and freeze all other weights \\
            \If{$t < N_k$}{
            Train $\theta_t$ by optimizing $L_t$ \tcp*{\cref{eq:loss_i}}
            }
            \Else{
            Train $\theta_t$ by optimizing $\| \hat{q}_{k_{N_{k}}} - q \|_2^2$
            }
        }
        Set all weights as trainable \\
        Train all weights by optimizing $\| \hat{q}_{k_{N_{k}}} - q \|_2^2$ \\
        \KwResult{Trained neural network parameters $\{\theta_1, ..., \theta_{N_{k}} \}$.}
        \caption{Training Procedure}
        \label{alg:warm_start_training}
    \end{algorithm}
    

%% file: revised_appendix_visualize_preds.tex
\section{Additional empirical results}
\label{sec:appendix_extra_plots}
\begin{landscape}
    \begin{figure}[!ht]
        \centering
        \includegraphics[width=\linewidth]{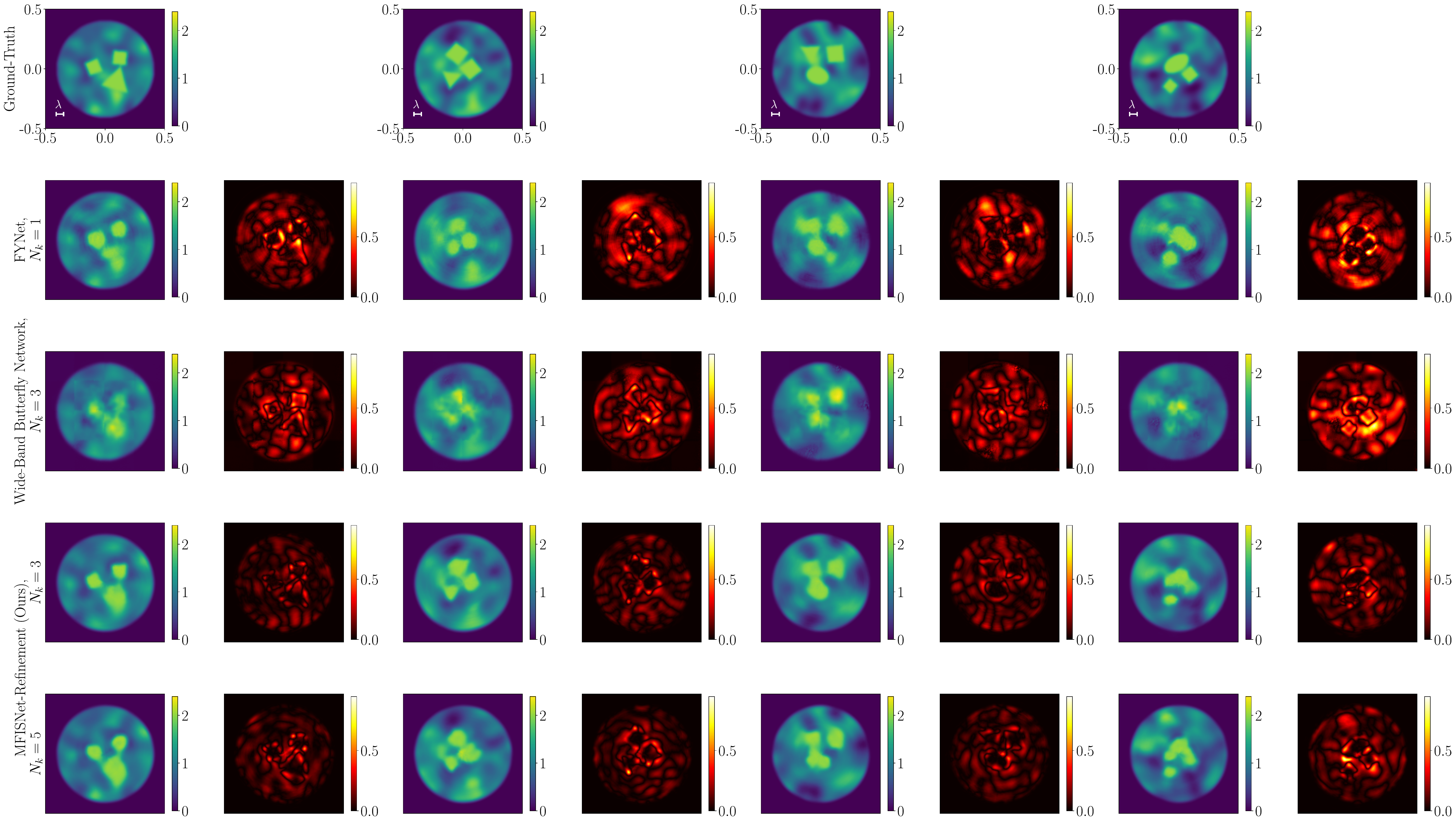}
        \caption{Sample predictions from four randomly-selected test samples. The first row shows the ground-truth scattering potential; in this plot we show the wavelength corresponding to the maximum frequency $k=32\pi$.
        }
        \label{fig:extra_preds_vis}
    \end{figure}
\end{landscape}

\begin{landscape}
    \begin{figure}[!ht]
        \centering
        \includegraphics[width=\linewidth]{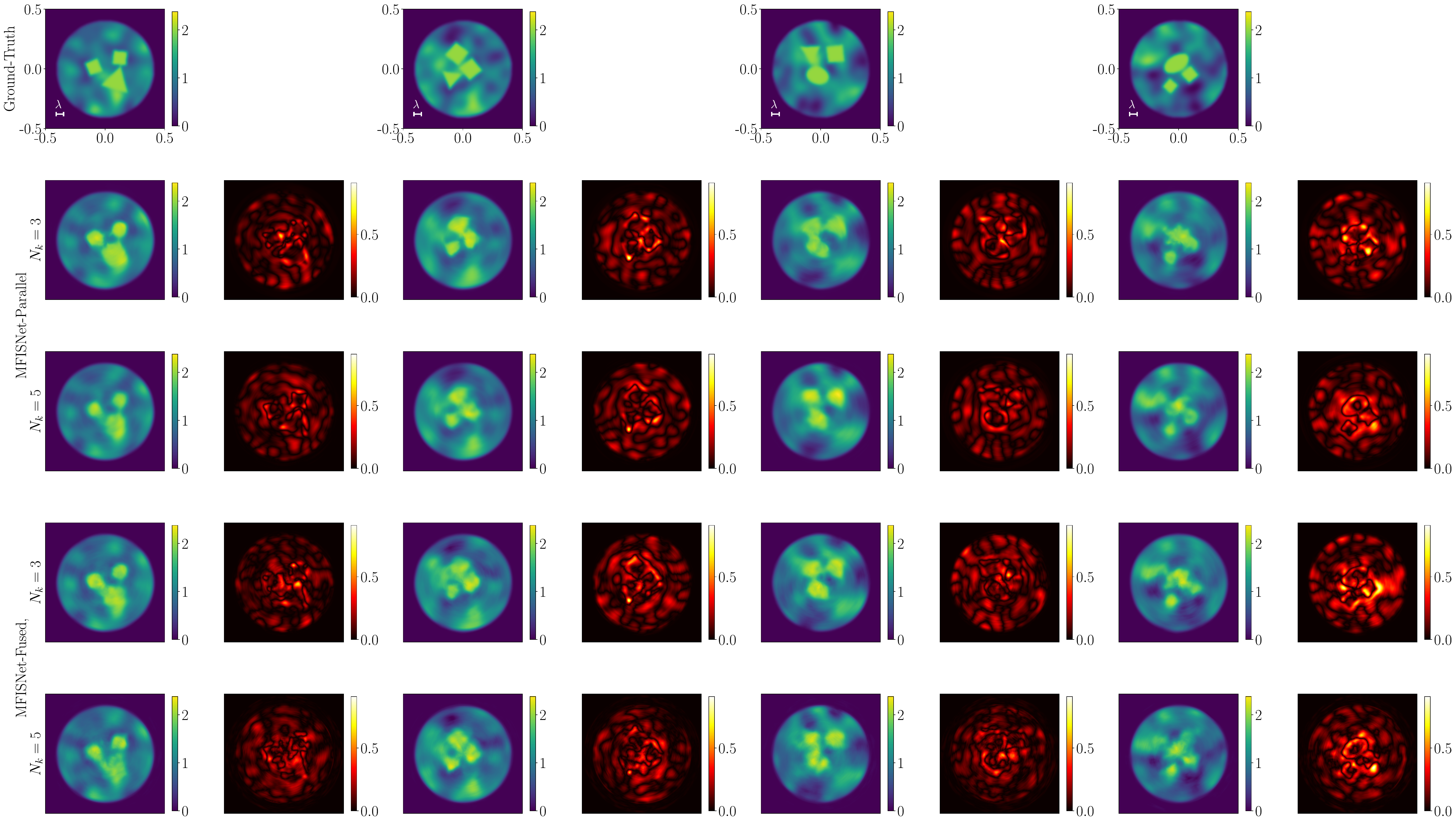}
        \caption{Sample predictions from four randomly-selected test samples. The first row shows the ground-truth scattering potential; in this plot we show the wavelength corresponding to the maximum frequency $k=32\pi$.
        }
        \label{fig:wide_parallel_vis}
    \end{figure}
\end{landscape}

In this section, we illustrate the predictions generated by the different models used in our
experiments on randomly-selected test samples from our dataset. For each prediction,
we include the associated error plot. In ~\cref{fig:extra_preds_vis}, we provide more samples comparing FYNet, Wide-Band Butterfly Network, and \ournet. In ~\cref{fig:wide_parallel_vis}, we show sample outputs from \wideinputnet~and \parallelnet.

